\documentclass[10pt,aps,prb,twocolumn,longbibliography,superscriptaddress,floatfix]{revtex4-2}

\usepackage{graphicx}% Include figure files
\usepackage{mathtools}
\usepackage{bm}% bold math
\usepackage{braket}
\renewcommand\vec\bm  % Bold vector instead of arrow

\usepackage{microtype}
\usepackage[acronym]{glossaries}
\usepackage{hyperref}% add hypertext capabilities
\usepackage{cleveref}
\usepackage{amsfonts}
\usepackage{xcolor}
\usepackage{longtable}

\begin{document}

\title{Origin of $A$-type antiferromagnetism and chiral split magnons in altermagnetic $\alpha$-MnTe}

\author{Mojtaba Alaei}
\affiliation{Materials Discovery Laboratory, Skolkovo Institute of Science and Technology, Bolshoy Boulevard 30, Building 1, Moscow 121205, Russia}
\affiliation{Department of Physics, Isfahan University of Technology, Isfahan 84156-83111, Iran}

\author{Pawel Sobieszczyk} 
\affiliation{Institute of Nuclear Physics Polish Academy of Sciences, Radzikowskiego 152, PL-31-342 Krakow, Poland}
\author{Andrzej Ptok} 
\affiliation{Institute of Nuclear Physics Polish Academy of Sciences, Radzikowskiego 152, 31-342 Krakow, Poland}
\author{Nafise Rezaei}
\affiliation{Materials Discovery Laboratory, Skolkovo Institute of Science and Technology, Bolshoy Boulevard 30, Building 1, Moscow 121205, Russia}
\author{Artem R. Oganov}
\affiliation{Materials Discovery Laboratory, Skolkovo Institute of Science and Technology, Bolshoy Boulevard 30, Building 1, Moscow 121205, Russia}
\author{Alireza Qaiumzadeh}
\affiliation{Center for Quantum Spintronics, Department of Physics, Norwegian University of Science and Technology, NO-7491 Trondheim, Norway}

\date{\today}

\begin{abstract}
The origin of the $A$-type antiferromagnetic ordering, characterized by ferromagnetic layers coupling antiferromagnetically, in the prototype semiconductor altermagnet $\alpha$-MnTe has been a topic of ongoing debate.
Experimentally, $\alpha$-MnTe exhibits an in-plane ferromagnetic exchange interaction, whereas previous \emph{ab initio} calculations predicted an antiferromagnetic interaction. In this paper, we resolve this discrepancy by considering an expanded set of magnetic configurations, which reveals a ferromagnetic in-plane exchange interaction in agreement with experimental findings. Additionally, we demonstrate that the 10th nearest-neighbor exchange interaction is directionally dependent, inducing a nonrelativistic chiral splitting in the magnon bands, as recently observed experimentally. We further show that applying a compressive strain may significantly enhance both nonrelativistic spin and chiral magnon splittings. The strain can also change the sign of the in-plane exchange interaction. Computing magnetic susceptibility, we show that strain enhances the N{\'e}el temperature, significantly. Our results highlight the critical importance of convergence in the number of magnetic configurations for spin interactions in antiferromagnetic materials.
\end{abstract}

\maketitle
\emph{Introduction--}
Antiferromagnetic (AFM) interactions give rise to a diverse range of AFM classes, ranging from different collinear structures to various exotic noncollinear and frustrated structures~\cite{RevModPhys.90.015005,Parkin,dal2024antiferromagnetic}.
A class of collinear AFM systems, in which combined inversion or translation and time-reversal symmetry ($\mathcal{IT}$ or $t\mathcal{T}$) is broken while combined crystal-rotation and time-reversal symmetry are retained, has recently been identified in various AFM materials, both theoretically ~\cite{noda2016momentum,naka2019spin,doi:10.7566/JPSJ.88.123702,PhysRevB.99.184432,vsmejkal2020crystal,PhysRevB.102.014422,PhysRevMaterials.5.014409,doi:10.1021/acs.jpcc.1c02653,mazin2021prediction,PhysRevLett.126.127701,Ma_2021,PhysRevX.12.040501,PhysRevX.12.011028,PhysRevX.12.031042,PhysRevB.102.014422,PhysRevLett.131.256703,Zunger-review,yuan2023uncovering,2024ApPhL.124c0503Y,PhysRevX.14.011019,S_dequist_2024,Cheong_2024,KIMEL2024172039,https://doi.org/10.1002/adfm.202409327,
jungwirth2024,jungwirth2024supefluid3healtermagnets} and experimentally~\cite{Krempasky2024,PhysRevLett.132.036702,PhysRevB.109.115102,doi:10.1126/sciadv.adj4883,Reimers_2024,  https://doi.org/10.1002/advs.202406529,ding2024largebandsplittinggwavetype}. These properties lead to the preservation of band degeneracy in the center of the Brillouin zone while removing the Kramers degeneracy in certain regions of the magnetic Brillouin zone. This class of \emph{nonrelativistic} spin-split collinear AFM materials has been termed altermagnetism. Therefore, in altermagnets, despite the absence of net magnetization, the electronic (magnonic) band structures display direction-dependent spin (chiral) splitting. Altermagnets generally belong to a broader class of nonrelativistic spin-split collinear AFM systems, which also include AFM half-metals, where the band degeneracy can even be broken at the center of the Brillouin zone  \cite{PhysRevLett.133.216701,doi:10.1073/pnas.1715465115,PhysRevLett.74.1171,osti_4642614}.

\begin{figure}[tb]
    \centering
    \includegraphics[width=\linewidth]{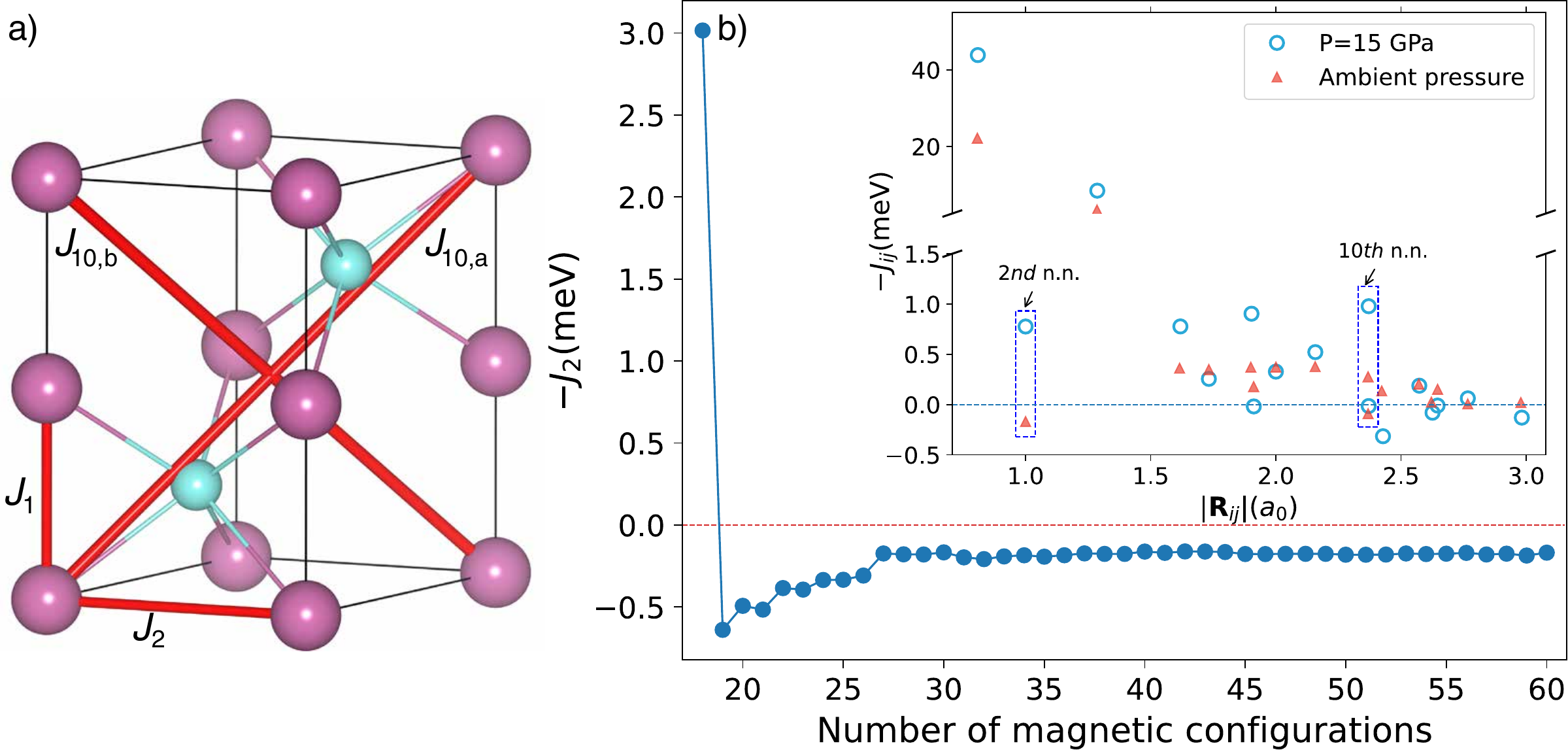}
    \caption{(a) Crystal structure of $\alpha$-MnTe. We present several important spin exchange interactions: the 10th n.n. interactions ($J_{10,a}$ and $J_{10,b}$) that lead to altermagnetism, as well as 1st. n.n. $J_1$ and 2nd n.n. $J_2$ that lead to $A$-type AFM order. Large purple spheres represent Mn atoms and small cyan spheres represent Te atoms.
    (b) 2nd n.n. normalized Heisenberg exchange interaction $J_2$, plotted against the number of magnetic configurations, starting from 18. Initially, $J_2$ is negative (AFM), but it transitions to positive (FM) and stabilizes as the number of configurations increases.
    The inset of (b) represents the normalized Heisenberg exchange interactions at ambient pressure $P=0$ and $P=15$ GPa.}
    \label{fig:MnTe_Js}
\end{figure}

$\alpha$-MnTe is a near-room-temperature, centrosymmetric correlated AFM semiconductor with a hexagonal NiAs-type crystal structure~\cite{Kriegner_2016,bossini2020exchange,PhysRevMaterials.8.104407}, recently identified experimentally as a candidate for $g$-wave altermagnetism~\cite{Krempasky2024,PhysRevLett.132.036702,PhysRevLett.132.176701,PhysRevB.109.115102,Chilcote_2024}. This material exhibits pronounced magnetostrictive and piezomagnetic properties~\cite{https://doi.org/10.1002/adfm.202305247,PhysRevMaterials.8.L041402}. 
The nearest-neighbor (n.n.) Mn-Mn bond is along the interlayer direction, while the 2nd n.n. is the in-plane bond; see Fig.~\ref{fig:MnTe_Js}(a). Experimentally, it was shown that the system has an $A$-type AFM structure \cite{PhysRevB.73.104403,Kriegner_2016,PhysRevB.96.214418,PhysRevLett.133.156702}, with an in-plane ferromagnetic (FM) exchange interaction $J_2 > 0$~\cite{PhysRevB.73.104403,PhysRevLett.133.156702}. In contrast, density functional theory (DFT) calculations found an AFM exchange interaction $J_2 < 0$ \cite{PhysRevB.107.L100418,PhysRevMaterials.3.025403,https://doi.org/10.1002/apxr.202300050},
which can induce magnetic frustrations. However, they proposed that stabilization of the $A$-type AFM order in this material relies on the presence of a strong 3rd n.n. interlayer AFM exchange interaction $J_3 < 0$. 

In another achievement, a recent inelastic neutron scattering experiment has measured chiral splitting of the magnon bands in this material \cite{PhysRevLett.133.156702}. Thus, $\alpha$ -MnTe shows altermagnetic properties in both electronic and magnonic band structures.

In this paper, using \emph{ab initio} calculations to compute the total energy, we resolve the previous discrepancy between DFT and experimental results. Our main finding, summarized in Fig.~\ref{fig:MnTe_Js}(b), shows that to find the correct sign for the in-plane exchange coupling $J_2$, one needs to take into account the sufficient number of magnetic configurations. Therefore, $J_2$ is responsible for $A$-type AFM order in this material. In addition, we demonstrate that the 10th n.n. spin exchange interaction  exhibits direction dependence, resulting in a nonrelativistic chiral splitting of the magnon dispersion in this material. Furthermore, we demonstrate that applying compressive strain enhances both spin-split electronic bands and chiral-split magnon bands. By combining DFT and atomistic spin dynamics simulations, we demonstrate that applied pressure reverses the sign of \( J_2 \); however, the magnetic ground state retains its collinear $A$-type AFM order, providing a novel avenue for controlling magnetic interactions.

\emph{Computational methods--}
To calculate the electronic band structures, the total energy, and consequently spin interactions, we use the projected augmented wave (PAW) method, as implemented in the Vienna \emph{ab initio} simulation package (VASP)~\cite{PhysRevB.59.1758}. The plane wave expansion employs a cutoff energy of 550 eV. We apply the generalized gradient approximation (GGA) developed by Perdew, Burke, and Ernzerhof (PBE) to account for the electron exchange-correlation energy~\cite{PBE}. Furthermore, we include a Hubbard correction of $U = 4$ eV, as estimated in Ref.~\cite{Mosleh2023}, to improve the treatment of electron-electron interactions.

To compute Heisenberg exchange interactions $\Tilde{J}_{ij}$, between localized spins $\mathbf{S}_{i}=S \hat{\mathbf{S}}_{i}$, where $S$ is the spin length and $\hat{\mathbf{S}}_{i}$ denotes the spin direction; we employ an implicit approach, named total energy mapping, by fitting a classical Heisenberg model $\mathcal{H}=-\sum_{i<j}J_{ij}\hat{\mathbf{S}}_{i}\cdot\hat{\mathbf{S}}_{j}$, where  ${J}_{ij}=S^2\Tilde{J}_{ij}$ are normalized Heisenberg exchange interactions, to the total energy derived from calculations of the electronic structure across numerous magnetic configurations~\cite{, Alaei_2023, Mosleh2023, RevModPhys.95.035004}. For this, we compute exchange interactions up to the 16th n.n., allowing us to capture the chiral magnon band-splitting accurately. To optimize computational efficiency, we select the minimal supercell that effectively captures all relevant exchange interactions. For this purpose, we apply the SUPERHEX method~\cite{superhex}, recently introduced by some of us. Using SUPERHEX, we obtain a supercell with just 34 Mn atoms. In contrast, a conventional approach would require a much larger supercell of $5\times5\times4$, which contains 200 Mn atoms. Our approach has thus enabled a speedup of two to three orders of magnitude.

At ambient pressure, we use approximately 60 unique magnetic configurations and apply a least-squares fitting to the Heisenberg Hamiltonian model. However, under a compressive pressure of 15 GPa, achieving convergence in exchange interactions requires about 120 magnetic configurations. In our DFT calculations, we neglect relativistic spin-orbit coupling effects to concentrate on the primary exchange interactions.

\begin{figure}[tb]
    \centering
    \includegraphics[width=\linewidth]{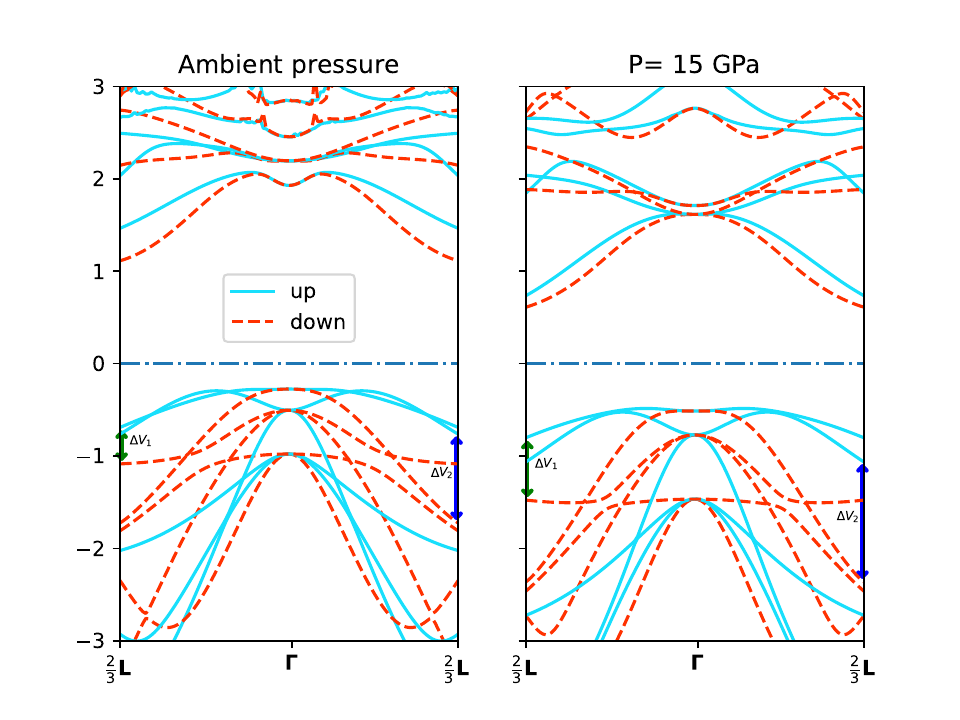}
    \caption{Electronic band structure of altermagnetic semiconductor $\alpha$-MnTe with $g$-wave symmetry of spin polarization
in momentum space. The left and right panels show the spin-resolved band structure of bulk $\alpha$-MnTe at ambient pressure and $15$~GPa, respectively. 
    $\Delta V_1$ and $\Delta V_2$ represent the spin subband splitting of the first and second valence bands, respectively.}
    \label{fig:bands_vs_P}
\end{figure}

\emph{Spin-resolved electronic band structure--}
In altermagnets, Kramers degeneracy is lifted, leading to a momentum-dependent splitting of spin subbands in the electronic band structure. The left and right panels in Fig.~\ref{fig:bands_vs_P} present the spin-resolved electronic band structure of the bulk $\alpha$-MnTe at ambient pressure and $15$~GPa, respectively, along the L-$\Gamma$ path in the Brillouin zone, where the material exhibits the largest spin splitting. We measure the nonrelativistic spin subband splitting at the $2/3$L k-point to compare its behavior under different pressures. $\Delta V_1$ represents the spin subband splitting of the first valence band, while $\Delta V_2$ corresponds to the second valence band; see Fig.~\ref{fig:bands_vs_P}. We found a large spin split of $\Delta V_1 = 0.39$ eV and $\Delta V_2 = 0.96$ eV under ambient conditions, consistent with recent experiments~\cite{PhysRevLett.132.036702,Krempasky2024}.
On the other hand, at $15$~GPa, $\Delta V_1$ increases to $0.67$~eV and $\Delta V_2$ to $1.3$~eV. This corresponds to about a 70\% increase in spin splitting of the first valence band and a 35\% increase for the second valence band when the system is under a pressure of $15$~GPa. We conclude that pressure enhances the nonrelativistic spin splitting in this material. We note that the NiAs-type structure of $\alpha$-MnTe undergoes a phase transition to a MnP-type structure under a higher pressure of 24 GPa~\cite{MnTe-phase}.

\emph{Normalized Heisenberg exchange interactions--}
In the inset of Fig.~\ref{fig:MnTe_Js}(b), we present the normalized Heisenberg exchange interactions, exchange spin interaction multiplied by $S^2$, as a function of distance, derived from DFT calculations of the total energy at ambient pressure and $15$~GPa. We present results up to the 16th n.n. Heisenberg exchange interactions, $J_1$-$J_{16}$, see also Table~\ref{tableI}. 
It is clear that increasing the pressure strengthens the Heisenberg exchange interactions for most exchange couplings. This enhancement can be attributed to the reduction in bond lengths, which increases the overlap of electron wave functions and consequently increases electron hopping.

Under ambient conditions, for the 2nd n.n., in-plane exchange interaction, our results predict an FM-type interaction, consistent with two experimental findings~\cite{PhysRevLett.133.156702,PhysRevB.73.104403}, while previous \emph{ab initio} studies report an AFM-type interaction~\cite{PhysRevB.107.L100418,PhysRevMaterials.3.025403,https://doi.org/10.1002/apxr.202300050}. Thus, the magnetic ground state is an $A$-type AFM as it was shown in experiments.

It should be noted that the $3d$-orbital occupancy of the Mn$^{2+}$ ions in $\alpha$-MnTe is approximately $5.3$, indicating that it is not exactly at half-filling. This slight deviation from half-filling can facilitate FM exchange interactions between in-plane Mn ions because the electron configuration allows for some degree of spin alignment.

We argue that this discrepancy between experiments and previous DFT calculations arises from using an insufficient number of magnetic configurations to map the \emph{ab initio} total energy to the Heisenberg Hamiltonian. Figure~\ref{fig:MnTe_Js}(b) clearly shows that relying on only 18 magnetic configurations also results in an incorrect AFM-type interaction for \( J_2 \) in our calculation. However, as we increase the number of magnetic configurations, the sign of the exchange interaction switches to the correct FM-type interaction, highlighting the importance of verifying the convergence of exchange interactions with respect to the number of magnetic configurations in complex AFM systems. 
Although the sign of $J_2$ in our calculations is in agreement with the experimental measurements, its amplitude is smaller~\cite{PhysRevLett.133.156702,PhysRevB.73.104403}. We believe that this may arise from the limited number of Heisenberg exchange interactions chosen for fitting the experimental data.
In addition, Fig.~\ref{fig:MnTe_Js}(b) shows that the sign of $J_2$ changes with pressure. Applying the Monte Carlo solver to atomistic spin dynamics~\cite{vampire}, we find that the magnetic ground state remains a collinear $A$-type AFM due to a large $J_3$ even though the sign of $J_2$ changes under applied pressure, see the Supplemental Material (SM)~\cite{SM}.

As we already mentioned, $\alpha$-MnTe is a prototype of altermagnetic systems. Partially lifting the band degeneracy in the electronic and magnonic band structure of this material has been experimentally reported very recently~\cite{PhysRevLett.132.036702,PhysRevLett.133.156702}.
Due to its crystal symmetry, then we expect some Heisenberg exchange interactions will exhibit two distinct interaction amplitudes in different directions. When considering exchanges up to the 16th n.n. in $\alpha$-MnTe, we find that the 10th n.n. consists of two values, which we label as $J_{10,a}$ and $J_{10,b}$. These two exchange interactions are illustrated in Fig.~\ref{fig:MnTe_Js}(a). In the figure, it is evident that the $J_{10,a}$ interaction is stronger than the $J_{10,b}$ interaction, due to the presence of Mn$-$Te bond connections in the $J_{10,a}$ interaction. 
The computed exchange interactions show that $J_{10,a}$ is stronger
than $J_{10,b}$, see the inset of Fig.~\ref{fig:MnTe_Js}(b). Under pressure, the difference between $J_{10,a}$ and $J_{10,b}$ increases, with $J_{10,a}$ changing from $-0.28$~meV at ambient pressure to $-0.98$~meV at $15$~GPa, resulting in an enhancement of the AFM magnon band splitting.

Consistent with a very recent experiment \cite{PhysRevLett.133.156702}, our calculations indicate a direction-dependent spin exchange interaction of $J_{10}$, i.e., $|J_{10,a} - J_{10,b}| \neq 0$. Such a direction dependence of exchange interactions, arising from the crystal geometry, leads to the nonrelativistic chiral split of magnon bands in antiferromagnets; see the magnon dispersions in the following. 
Furthermore, in agreement with the experiment finding, we find that $J_{10,a}$ exhibits an AFM-type sign while $J_{10,b}$ shows an FM-type sign. We find that while applying the pressure cannot change the sign of them, their difference is enhanced and thus the chiral splitting of magnon bands is increased.
We are not aware of any previous \emph{ab initio} study of direction-dependent Heisenberg spin exchange interactions in $\alpha$-MnTe that leads to the chiral splitting of magnon bands.

\begin{table}[h!] 
    \centering
    \caption{Comparison of normalized Heisenberg exchange interactions, $J_n=S^2 \tilde{J}_n$, with \( n = 1, \dots, 16 \); computed in the present study at ambient pressure (\( P = 0 \)) and high pressure (\( P = 15 \) GPa), with recent experimental results, obtained by fitting the measured magnon dispersion to a simplified spin Hamiltonian model incorporating only five exchange interactions~\cite{PhysRevLett.133.156702}.}
    \begin{tabular}{cccccc}
        \hline
        $J_n (\text{meV})$ & $P=0$ & $P= 15$ GPa & \text{Exp.~\cite{PhysRevLett.133.156702}} \\
        \hline
        \( J_1 \) & -22.1816 & -43.8556 & -24.94  \\
        \( J_2 \) & 0.1686 & -0.7762 & 0.75  \\
        \( J_3 \) & -3.4239 & -8.4657 & -2.95  \\
        \( J_4 \) & -0.3620 & -0.7770 & -- \\
        \( J_5 \) & -0.3485 & -0.2554 & -- \\
        \( J_6 \) & -0.3714 & -0.9036 & -- \\
        \( J_7 \) & -0.1795 & 0.0185 & -- \\
        \( J_8 \) & -0.3734 & -0.3274 & -- \\
        \( J_9 \) & -0.3770 & -0.5216 & -- \\
        \( J_{10,a} \) & -0.2772 & -0.9780 & -0.425 \\
        \( J_{10,b} \) & 0.0907 & 0.0147 & 0.1381 \\
        \( J_{11} \) & -0.1361 & 0.3152 & -- \\
        \( J_{12} \) & -0.2042 & -0.1877 & -- \\
        \( J_{13} \) & -0.0242 & 0.0810 & --  \\
        \( J_{14} \) & -0.1523 & 0.0100 & -- \\
        \( J_{15} \) & -0.0062 & -0.0617 & -- \\
        \( J_{16} \) & -0.0203 & 0.1297 & -- \\
        \hline
    \end{tabular}
    \label{tableI}
\end{table}

\emph{Magnon dispersion and magnetic susceptibility--}
Finding normalized Heisenberg exchange interactions, see Table~\ref{tableI}, we are able to compute magnon dispersion under both ambient conditions and under pressure as well as magnetic ground state and magnetic susceptibility. We use the following minimal spin Hamiltonian to describe the spin interactions in $\alpha$-MnTe,
\begin{equation}
\label{eqn:Hamiltonian}
\mathcal{H}=-\sum_{i<j}J_{ij}\hat{\mathbf{S}}_{i}\cdot\hat{\mathbf{S}}_{j}-K\sum_{i}(\hat{\mathbf{S}}_i\cdot\hat{\mathbf{e}}_i)^{2}-\mu_{s}h_{0}\sum_{i}\hat{\mathbf{b}}\cdot\hat{\mathbf{S}}_{i},
\end{equation}
where $K>0$ is the single-ion uniaxial easy-axis magnetic anisotropy constant, $\hat{\mathbf{e}}_i$ is the magnetic anisotropy direction, and $\mu_{s}$ is the atomic magnetic moment. We assume a Zeeman-like interaction between localized magnetic moments and an external magnetic field with amplitude $h_{0}$ along the $\hat{\mathbf{b}}$ direction.
The crystalline magnetic anisotropy, which arises from spin--orbit coupling, is very weak in $\alpha$-MnTe. Due to its small magnitude, it cannot be reliably determined either experimentally or through \emph{ab initio} calculations. For our analysis, we adopt a value of $K = 6.25$~meV \cite{PhysRevB.73.104403,PhysRevB.107.L100418}. In collinear uniaxial AFM systems, magnons can have two chiral $\alpha$ and $\beta$ eigenmodes, $\mathcal{H}= \sum_{k} (\omega_k^{\alpha} \alpha_k^\dag \alpha_k +\omega_k^{\beta} \beta_k^\dag \beta_k)$~\cite{10.1063/1.5109132}. In conventional $\mathcal{IT}$ or $t\mathcal{T}$ symmetric AFM systems, these two chiral modes are usually degenerate $\omega_k^{\alpha}=\omega_k^{\beta}$. However, in altermagnets, the chiral magnon bands split~\cite{PhysRevLett.131.256703} as was shown in a recent experiment on $\alpha$-MnTe~\cite{PhysRevLett.133.156702}. 

\begin{figure}[tb]
    \centering
    \includegraphics[width=\linewidth]{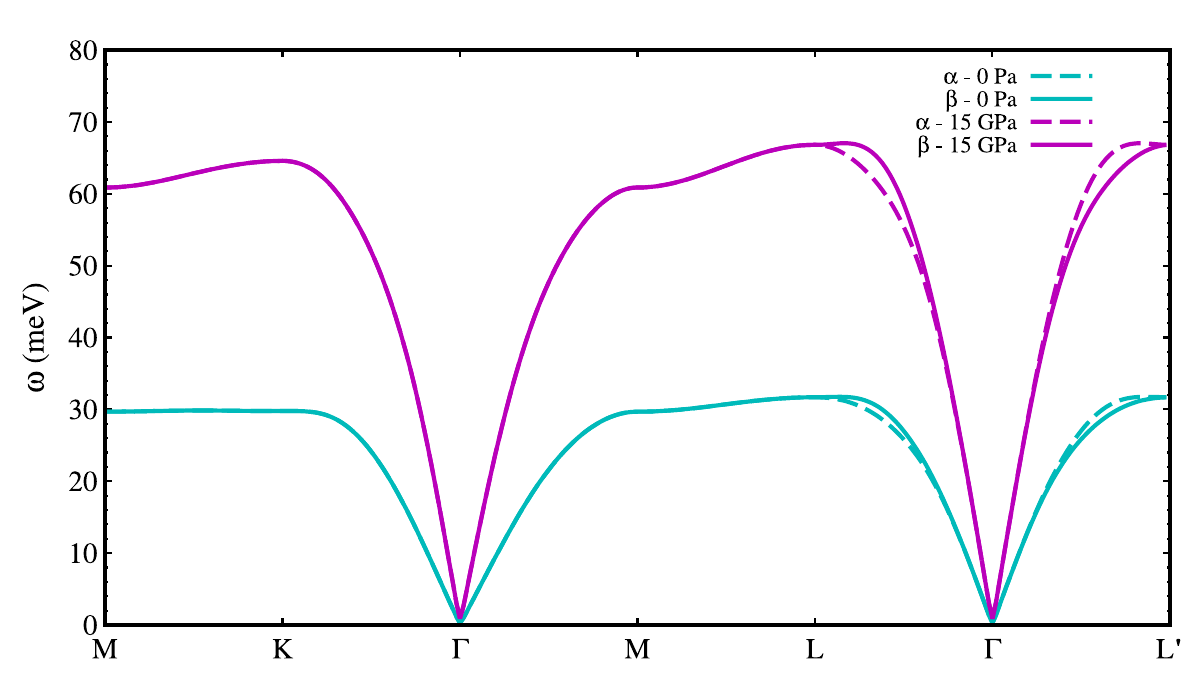}
    \caption{Magnon dispersion relation of $\alpha$-MnTe in the ambient conditions (green lines) and under a compressive pressure of 15 GPa  (purple lines). In the absence of magnetic field, chiral $\alpha$ (dashed lines) and $\beta$ (solid lines) magnon subbands are degenerate around the $\Gamma$ symmetry point, while the chiral degeneracy is lifted along the L-$\Gamma$-L symmetry path. See the SM for magnon dispersions in the presence of a magnetic field~\cite{SM}.}
    \label{fig:dispersionMnTe}
\end{figure}

Figure~\ref{fig:dispersionMnTe} shows the magnon dispersion of $\alpha$-MnTe in the absence and presence of pressure. Low energy magnons around the $\Gamma$ symmetry point are degenerate, while along the L-$\Gamma$-L symmetry path this degeneracy is broken. The pressure enhances the band splitting and also the magnon bandwidth. The spin-flop magnetic field in AFM systems is proportional to $\sqrt{J_1 K}$. Using dispersion relation calculations, we find a spin-flop magnetic field of $h_{\rm{sf}} \approx 5.3$ T in the absence of pressure, while in the presence of the pressure it enhances to $h_{\rm{sf}} \approx 7.6$~T. In the SM~\cite{SM}, we show the magnetic field dependence of the chiral splitting.

In SM \cite{SM}, we have compared the chiral splitting of the magnon bands, derived from our \emph{ab initio} calculations, with the experimental data available in Ref. \cite{PhysRevLett.133.156702} and found that the maximum chiral splitting according to our calculations is 1.5 times larger than the experimental data. The difference between our findings and the experimental data can have different origins: (i) The experimental data were derived by fitting to an exchange Hamiltonian that includes only five exchange terms, and (ii) DFT+$U$ calculations are known to underestimate exchange interactions~\cite{Mosleh2023}. 
Further experimental investigations, such as magnetic field-dependent and temperature-dependent studies, are essential to achieve a comprehensive characterization of this material.

\begin{figure}[t]
    \centering
    \includegraphics[width=\linewidth]{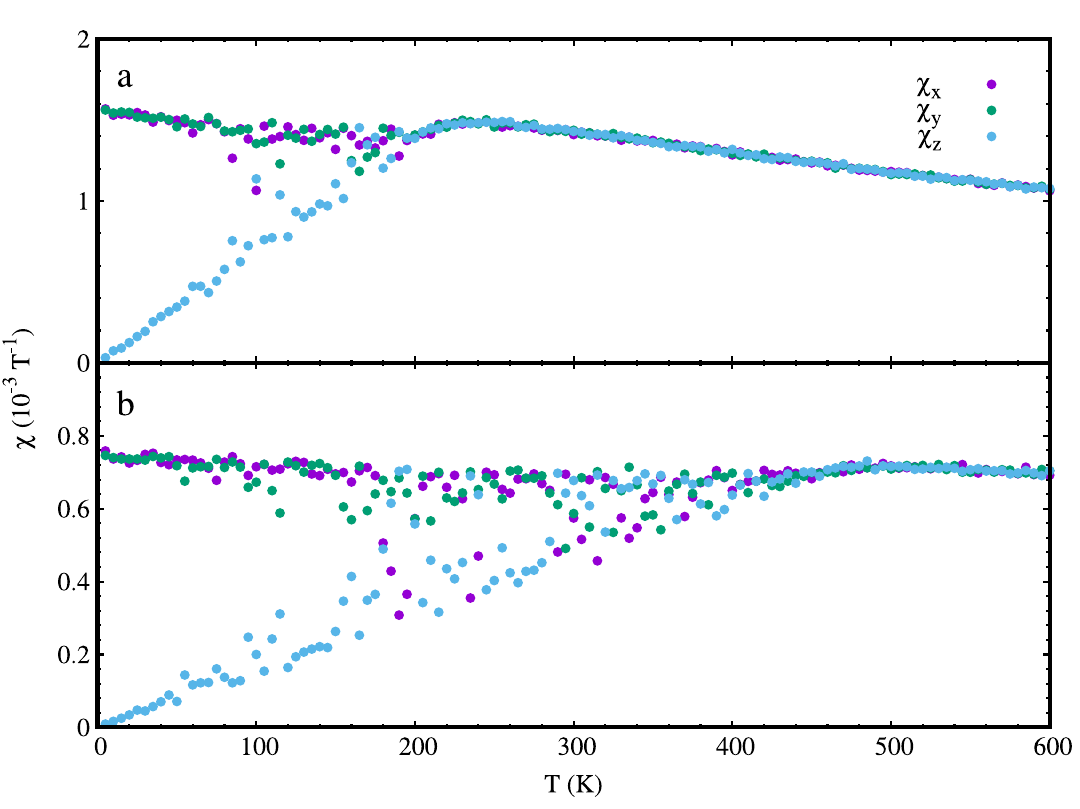}
    \caption{Directional magnetic susceptibilities in the (a) absence and (b) presence of pressure . In our atomistic simulations, we model a cubic system with dimensions of \( 10 \, \text{nm} \times 10 \, \text{nm} \times 10 \, \text{nm} \), with periodic boundary conditions. }
    \label{fig:susceptlibilityP0Pa}
\end{figure}

In addition, we compute the magnetic susceptibility of the system using Monte Carlo calculations in Fig.~\ref{fig:susceptlibilityP0Pa}. The N{\'e}el temperature $T_{\rm{N}}$ can be read out from the maximum of the longitudinal magnetic susceptibility. Within our calculations, we find $T_{\rm{N}} \approx 250$ K in the absence of pressure and $T_{\rm{N}} \approx 500$~K under the pressure. Our calculations yield a lower N{\'e}el temperature compared to the experimental values, which are reported as $267$~K for the thin film~\cite{PhysRevLett.132.036702} and $307$~K for the bulk material~\cite{PhysRevLett.133.156702,PhysRevB.109.214434}. This discrepancy is probably due to an underestimation of the exchange interaction parameter \(J_1\) in our calculations.

Finally, from inverse magnetic susceptibility calculations, see SM~\cite{SM}, we obtain the AFM Curie--Weiss temperature $\Theta$ via phenomenological Curie--Weiss law $\chi_z^{-1} \propto (T + \Theta)$ \cite{Mugiraneza_2022}. Under ambient conditions, we find $|\Theta| \approx 620$~K, which is slightly higher than the previous experimental values of $585$~K reported in Refs.~\cite{doi:10.1021/j100822a006,doi:10.1143/JPSJ.18.356}. However, under applied pressure, $|\Theta|$ increases significantly to around 3100 K. 
These results are in agreement with the typical frustration index $|\Theta|/T_{\rm{N}}\propto 2-5$ of unfrustrated 3$d$ transition metals~\cite{Mugiraneza_2022}.
 
\emph{Concluding remarks--}
$\alpha$-MnTe serves as a prototypical altermagnetic semiconductor, distinguished by its robust piezomagnetic properties. These characteristics make it a compelling candidate for advanced spintronics applications. In this study, we address two open questions: the origin of $A$-type AFM order in this material and the chiral split of magnon bands. By resolving the in-plane Heisenberg exchange interaction $J_2$ discrepancy between the experimental findings and prior DFT calculations, we highlight the critical role of magnetic configurations in accurately modeling complex antiferromagnets. Furthermore, we identify \( J_{10a(b)} \) as the primary driver behind the splitting of chiral magnons in $\alpha$-MnTe. Notably, applied pressure modulates both the sign and magnitude of the Heisenberg exchange interactions, enhancing spin polarization and chiral band splitting in both electronic and magnonic spectra. This work underscores the importance of a detailed spin interaction analysis in advancing the physics of altermagnetic materials for next-generation spintronics technologies.

\emph{Acknowledgment--}
This work has been supported by the Research Council of Norway through its Centers of Excellence funding scheme, Project No. 262633, ``QuSpin''. A.R.O. was supported by
the Russian Science Foundation under Grant No. 19-72-30043.

\bibliography{ref.bib}

\begin{thebibliography}{66}%
\makeatletter
\providecommand \@ifxundefined [1]{%
 \@ifx{#1\undefined}
}%
\providecommand \@ifnum [1]{%
 \ifnum #1\expandafter \@firstoftwo
 \else \expandafter \@secondoftwo
 \fi
}%
\providecommand \@ifx [1]{%
 \ifx #1\expandafter \@firstoftwo
 \else \expandafter \@secondoftwo
 \fi
}%
\providecommand \natexlab [1]{#1}%
\providecommand \enquote  [1]{``#1''}%
\providecommand \bibnamefont  [1]{#1}%
\providecommand \bibfnamefont [1]{#1}%
\providecommand \citenamefont [1]{#1}%
\providecommand \href@noop [0]{\@secondoftwo}%
\providecommand \href [0]{\begingroup \@sanitize@url \@href}%
\providecommand \@href[1]{\@@startlink{#1}\@@href}%
\providecommand \@@href[1]{\endgroup#1\@@endlink}%
\providecommand \@sanitize@url [0]{\catcode `\\12\catcode `\$12\catcode `\&12\catcode `\#12\catcode `\^12\catcode `\_12\catcode `\%12\relax}%
\providecommand \@@startlink[1]{}%
\providecommand \@@endlink[0]{}%
\providecommand \url  [0]{\begingroup\@sanitize@url \@url }%
\providecommand \@url [1]{\endgroup\@href {#1}{\urlprefix }}%
\providecommand \urlprefix  [0]{URL }%
\providecommand \Eprint [0]{\href }%
\providecommand \doibase [0]{https://doi.org/}%
\providecommand \selectlanguage [0]{\@gobble}%
\providecommand \bibinfo  [0]{\@secondoftwo}%
\providecommand \bibfield  [0]{\@secondoftwo}%
\providecommand \translation [1]{[#1]}%
\providecommand \BibitemOpen [0]{}%
\providecommand \bibitemStop [0]{}%
\providecommand \bibitemNoStop [0]{.\EOS\space}%
\providecommand \EOS [0]{\spacefactor3000\relax}%
\providecommand \BibitemShut  [1]{\csname bibitem#1\endcsname}%
\let\auto@bib@innerbib\@empty
%</preamble>
\bibitem [{\citenamefont {Baltz}\ \emph {et~al.}(2018)\citenamefont {Baltz}, \citenamefont {Manchon}, \citenamefont {Tsoi}, \citenamefont {Moriyama}, \citenamefont {Ono},\ and\ \citenamefont {Tserkovnyak}}]{RevModPhys.90.015005}%
  \BibitemOpen
  \bibfield  {author} {\bibinfo {author} {\bibfnamefont {V.}~\bibnamefont {Baltz}}, \bibinfo {author} {\bibfnamefont {A.}~\bibnamefont {Manchon}}, \bibinfo {author} {\bibfnamefont {M.}~\bibnamefont {Tsoi}}, \bibinfo {author} {\bibfnamefont {T.}~\bibnamefont {Moriyama}}, \bibinfo {author} {\bibfnamefont {T.}~\bibnamefont {Ono}},\ and\ \bibinfo {author} {\bibfnamefont {Y.}~\bibnamefont {Tserkovnyak}},\ }\bibfield  {title} {\bibinfo {title} {Antiferromagnetic spintronics},\ }\href {https://doi.org/10.1103/RevModPhys.90.015005} {\bibfield  {journal} {\bibinfo  {journal} {Rev. Mod. Phys.}\ }\textbf {\bibinfo {volume} {90}},\ \bibinfo {pages} {015005} (\bibinfo {year} {2018})}\BibitemShut {NoStop}%
\bibitem [{\citenamefont {Rimmler}\ \emph {et~al.}(2024)\citenamefont {Rimmler}, \citenamefont {Pal},\ and\ \citenamefont {Parkin}}]{Parkin}%
  \BibitemOpen
  \bibfield  {author} {\bibinfo {author} {\bibfnamefont {B.}~\bibnamefont {Rimmler}}, \bibinfo {author} {\bibfnamefont {B.}~\bibnamefont {Pal}},\ and\ \bibinfo {author} {\bibfnamefont {S.}~\bibnamefont {Parkin}},\ }\bibfield  {title} {\bibinfo {title} {Non-collinear antiferromagnetic spintronics},\ }\href {https://doi.org/10.1038/s41578-024-00706-w} {\bibfield  {journal} {\bibinfo  {journal} {Nat. Rev. Mater.}\ }\textbf {\bibinfo {volume} {10}},\ \bibinfo {pages} {109} (\bibinfo {year} {2024})}\BibitemShut {NoStop}%
\bibitem [{\citenamefont {Dal~Din}\ \emph {et~al.}(2024)\citenamefont {Dal~Din}, \citenamefont {Amin}, \citenamefont {Wadley},\ and\ \citenamefont {Edmonds}}]{dal2024antiferromagnetic}%
  \BibitemOpen
  \bibfield  {author} {\bibinfo {author} {\bibfnamefont {A.}~\bibnamefont {Dal~Din}}, \bibinfo {author} {\bibfnamefont {O.}~\bibnamefont {Amin}}, \bibinfo {author} {\bibfnamefont {P.}~\bibnamefont {Wadley}},\ and\ \bibinfo {author} {\bibfnamefont {K.}~\bibnamefont {Edmonds}},\ }\bibfield  {title} {\bibinfo {title} {Antiferromagnetic spintronics and beyond},\ }\href {https://doi.org/10.1038/s44306-024-00029-0} {\bibfield  {journal} {\bibinfo  {journal} {npj Spintronics}\ }\textbf {\bibinfo {volume} {2}},\ \bibinfo {pages} {25} (\bibinfo {year} {2024})}\BibitemShut {NoStop}%
\bibitem [{\citenamefont {Noda}\ \emph {et~al.}(2016)\citenamefont {Noda}, \citenamefont {Ohno},\ and\ \citenamefont {Nakamura}}]{noda2016momentum}%
  \BibitemOpen
  \bibfield  {author} {\bibinfo {author} {\bibfnamefont {Y.}~\bibnamefont {Noda}}, \bibinfo {author} {\bibfnamefont {K.}~\bibnamefont {Ohno}},\ and\ \bibinfo {author} {\bibfnamefont {S.}~\bibnamefont {Nakamura}},\ }\bibfield  {title} {\bibinfo {title} {{Momentum-dependent band spin splitting in semiconducting MnO$_2$: a density functional calculation}},\ }\href {https://doi.org/10.1039/C5CP07806G} {\bibfield  {journal} {\bibinfo  {journal} {Phys. Chem. Chem. Phys.}\ }\textbf {\bibinfo {volume} {18}},\ \bibinfo {pages} {13294} (\bibinfo {year} {2016})}\BibitemShut {NoStop}%
\bibitem [{\citenamefont {Naka}\ \emph {et~al.}(2019)\citenamefont {Naka}, \citenamefont {Hayami}, \citenamefont {Kusunose}, \citenamefont {Yanagi}, \citenamefont {Motome},\ and\ \citenamefont {Seo}}]{naka2019spin}%
  \BibitemOpen
  \bibfield  {author} {\bibinfo {author} {\bibfnamefont {M.}~\bibnamefont {Naka}}, \bibinfo {author} {\bibfnamefont {S.}~\bibnamefont {Hayami}}, \bibinfo {author} {\bibfnamefont {H.}~\bibnamefont {Kusunose}}, \bibinfo {author} {\bibfnamefont {Y.}~\bibnamefont {Yanagi}}, \bibinfo {author} {\bibfnamefont {Y.}~\bibnamefont {Motome}},\ and\ \bibinfo {author} {\bibfnamefont {H.}~\bibnamefont {Seo}},\ }\bibfield  {title} {\bibinfo {title} {Spin current generation in organic antiferromagnets},\ }\href {https://doi.org/10.1038/s41467-019-12229-y} {\bibfield  {journal} {\bibinfo  {journal} {Nat. Commun.}\ }\textbf {\bibinfo {volume} {10}},\ \bibinfo {pages} {4305} (\bibinfo {year} {2019})}\BibitemShut {NoStop}%
\bibitem [{\citenamefont {Hayami}\ \emph {et~al.}(2019)\citenamefont {Hayami}, \citenamefont {Yanagi},\ and\ \citenamefont {Kusunose}}]{doi:10.7566/JPSJ.88.123702}%
  \BibitemOpen
  \bibfield  {author} {\bibinfo {author} {\bibfnamefont {S.}~\bibnamefont {Hayami}}, \bibinfo {author} {\bibfnamefont {Y.}~\bibnamefont {Yanagi}},\ and\ \bibinfo {author} {\bibfnamefont {H.}~\bibnamefont {Kusunose}},\ }\bibfield  {title} {\bibinfo {title} {Momentum-dependent spin splitting by collinear antiferromagnetic ordering},\ }\href {https://doi.org/10.7566/JPSJ.88.123702} {\bibfield  {journal} {\bibinfo  {journal} {J. Phys. Soc. Jpn.}\ }\textbf {\bibinfo {volume} {88}},\ \bibinfo {pages} {123702} (\bibinfo {year} {2019})}\BibitemShut {NoStop}%
\bibitem [{\citenamefont {Ahn}\ \emph {et~al.}(2019)\citenamefont {Ahn}, \citenamefont {Hariki}, \citenamefont {Lee},\ and\ \citenamefont {Kune\ifmmode~\check{s}\else \v{s}\fi{}}}]{PhysRevB.99.184432}%
  \BibitemOpen
  \bibfield  {author} {\bibinfo {author} {\bibfnamefont {K.-H.}\ \bibnamefont {Ahn}}, \bibinfo {author} {\bibfnamefont {A.}~\bibnamefont {Hariki}}, \bibinfo {author} {\bibfnamefont {K.-W.}\ \bibnamefont {Lee}},\ and\ \bibinfo {author} {\bibfnamefont {J.}~\bibnamefont {Kune\ifmmode~\check{s}\else \v{s}\fi{}}},\ }\bibfield  {title} {\bibinfo {title} {{Antiferromagnetism in ${\mathrm{RuO}}_{2}$ as $d$-wave Pomeranchuk instability}},\ }\href {https://doi.org/10.1103/PhysRevB.99.184432} {\bibfield  {journal} {\bibinfo  {journal} {Phys. Rev. B}\ }\textbf {\bibinfo {volume} {99}},\ \bibinfo {pages} {184432} (\bibinfo {year} {2019})}\BibitemShut {NoStop}%
\bibitem [{\citenamefont {{\v{S}}mejkal}\ \emph {et~al.}(2020)\citenamefont {{\v{S}}mejkal}, \citenamefont {Gonz{\'a}lez-Hern{\'a}ndez}, \citenamefont {Jungwirth},\ and\ \citenamefont {Sinova}}]{vsmejkal2020crystal}%
  \BibitemOpen
  \bibfield  {author} {\bibinfo {author} {\bibfnamefont {L.}~\bibnamefont {{\v{S}}mejkal}}, \bibinfo {author} {\bibfnamefont {R.}~\bibnamefont {Gonz{\'a}lez-Hern{\'a}ndez}}, \bibinfo {author} {\bibfnamefont {T.}~\bibnamefont {Jungwirth}},\ and\ \bibinfo {author} {\bibfnamefont {J.}~\bibnamefont {Sinova}},\ }\bibfield  {title} {\bibinfo {title} {{Crystal time-reversal symmetry breaking and spontaneous Hall effect in collinear antiferromagnets}},\ }\href {https://doi.org/10.1126/sciadv.aaz88} {\bibfield  {journal} {\bibinfo  {journal} {Sci. Adv.}\ }\textbf {\bibinfo {volume} {6}},\ \bibinfo {pages} {eaaz8809} (\bibinfo {year} {2020})}\BibitemShut {NoStop}%
\bibitem [{\citenamefont {Yuan}\ \emph {et~al.}(2020)\citenamefont {Yuan}, \citenamefont {Wang}, \citenamefont {Luo}, \citenamefont {Rashba},\ and\ \citenamefont {Zunger}}]{PhysRevB.102.014422}%
  \BibitemOpen
  \bibfield  {author} {\bibinfo {author} {\bibfnamefont {L.-D.}\ \bibnamefont {Yuan}}, \bibinfo {author} {\bibfnamefont {Z.}~\bibnamefont {Wang}}, \bibinfo {author} {\bibfnamefont {J.-W.}\ \bibnamefont {Luo}}, \bibinfo {author} {\bibfnamefont {E.~I.}\ \bibnamefont {Rashba}},\ and\ \bibinfo {author} {\bibfnamefont {A.}~\bibnamefont {Zunger}},\ }\bibfield  {title} {\bibinfo {title} {{Giant momentum-dependent spin splitting in centrosymmetric low-$Z$ antiferromagnets}},\ }\href {https://doi.org/10.1103/PhysRevB.102.014422} {\bibfield  {journal} {\bibinfo  {journal} {Phys. Rev. B}\ }\textbf {\bibinfo {volume} {102}},\ \bibinfo {pages} {014422} (\bibinfo {year} {2020})}\BibitemShut {NoStop}%
\bibitem [{\citenamefont {Yuan}\ \emph {et~al.}(2021)\citenamefont {Yuan}, \citenamefont {Wang}, \citenamefont {Luo},\ and\ \citenamefont {Zunger}}]{PhysRevMaterials.5.014409}%
  \BibitemOpen
  \bibfield  {author} {\bibinfo {author} {\bibfnamefont {L.-D.}\ \bibnamefont {Yuan}}, \bibinfo {author} {\bibfnamefont {Z.}~\bibnamefont {Wang}}, \bibinfo {author} {\bibfnamefont {J.-W.}\ \bibnamefont {Luo}},\ and\ \bibinfo {author} {\bibfnamefont {A.}~\bibnamefont {Zunger}},\ }\bibfield  {title} {\bibinfo {title} {{Prediction of low-Z collinear and noncollinear antiferromagnetic compounds having momentum-dependent spin splitting even without spin-orbit coupling}},\ }\href {https://doi.org/10.1103/PhysRevMaterials.5.014409} {\bibfield  {journal} {\bibinfo  {journal} {Phys. Rev. Mater.}\ }\textbf {\bibinfo {volume} {5}},\ \bibinfo {pages} {014409} (\bibinfo {year} {2021})}\BibitemShut {NoStop}%
\bibitem [{\citenamefont {Egorov}\ \emph {et~al.}(2021)\citenamefont {Egorov}, \citenamefont {Litvin},\ and\ \citenamefont {Evarestov}}]{doi:10.1021/acs.jpcc.1c02653}%
  \BibitemOpen
  \bibfield  {author} {\bibinfo {author} {\bibfnamefont {S.~A.}\ \bibnamefont {Egorov}}, \bibinfo {author} {\bibfnamefont {D.~B.}\ \bibnamefont {Litvin}},\ and\ \bibinfo {author} {\bibfnamefont {R.~A.}\ \bibnamefont {Evarestov}},\ }\bibfield  {title} {\bibinfo {title} {{Antiferromagnetism-Induced Spin Splitting in Systems Described by Magnetic Layer Groups}},\ }\href {https://doi.org/10.1021/acs.jpcc.1c02653} {\bibfield  {journal} {\bibinfo  {journal} {J. Phys. Chem. C}\ }\textbf {\bibinfo {volume} {125}},\ \bibinfo {pages} {16147} (\bibinfo {year} {2021})}\BibitemShut {NoStop}%
\bibitem [{\citenamefont {Mazin}\ \emph {et~al.}(2021)\citenamefont {Mazin}, \citenamefont {Koepernik}, \citenamefont {Johannes}, \citenamefont {Gonz{\'a}lez-Hern{\'a}ndez},\ and\ \citenamefont {{\v{S}}mejkal}}]{mazin2021prediction}%
  \BibitemOpen
  \bibfield  {author} {\bibinfo {author} {\bibfnamefont {I.~I.}\ \bibnamefont {Mazin}}, \bibinfo {author} {\bibfnamefont {K.}~\bibnamefont {Koepernik}}, \bibinfo {author} {\bibfnamefont {M.~D.}\ \bibnamefont {Johannes}}, \bibinfo {author} {\bibfnamefont {R.}~\bibnamefont {Gonz{\'a}lez-Hern{\'a}ndez}},\ and\ \bibinfo {author} {\bibfnamefont {L.}~\bibnamefont {{\v{S}}mejkal}},\ }\bibfield  {title} {\bibinfo {title} {{Prediction of unconventional magnetism in doped FeSb$_2$}},\ }\href {https://doi.org/10.1073/pnas.2108924118} {\bibfield  {journal} {\bibinfo  {journal} {Proc. Natl. Acad. Sci. U.S.A.}\ }\textbf {\bibinfo {volume} {118}},\ \bibinfo {pages} {e2108924118} (\bibinfo {year} {2021})}\BibitemShut {NoStop}%
\bibitem [{\citenamefont {Gonz\'alez-Hern\'andez}\ \emph {et~al.}(2021)\citenamefont {Gonz\'alez-Hern\'andez}, \citenamefont {\ifmmode~\check{S}\else \v{S}\fi{}mejkal}, \citenamefont {V\'yborn\'y}, \citenamefont {Yahagi}, \citenamefont {Sinova}, \citenamefont {Jungwirth},\ and\ \citenamefont {\ifmmode~\check{Z}\else \v{Z}\fi{}elezn\'y}}]{PhysRevLett.126.127701}%
  \BibitemOpen
  \bibfield  {author} {\bibinfo {author} {\bibfnamefont {R.}~\bibnamefont {Gonz\'alez-Hern\'andez}}, \bibinfo {author} {\bibfnamefont {L.}~\bibnamefont {\ifmmode~\check{S}\else \v{S}\fi{}mejkal}}, \bibinfo {author} {\bibfnamefont {K.}~\bibnamefont {V\'yborn\'y}}, \bibinfo {author} {\bibfnamefont {Y.}~\bibnamefont {Yahagi}}, \bibinfo {author} {\bibfnamefont {J.}~\bibnamefont {Sinova}}, \bibinfo {author} {\bibfnamefont {T.}~\bibnamefont {Jungwirth}},\ and\ \bibinfo {author} {\bibfnamefont {J.}~\bibnamefont {\ifmmode~\check{Z}\else \v{Z}\fi{}elezn\'y}},\ }\bibfield  {title} {\bibinfo {title} {{Efficient Electrical Spin Splitter Based on Nonrelativistic Collinear Antiferromagnetism}},\ }\href {https://doi.org/10.1103/PhysRevLett.126.127701} {\bibfield  {journal} {\bibinfo  {journal} {Phys. Rev. Lett.}\ }\textbf {\bibinfo {volume} {126}},\ \bibinfo {pages} {127701} (\bibinfo {year} {2021})}\BibitemShut {NoStop}%
\bibitem [{\citenamefont {Ma}\ \emph {et~al.}(2021)\citenamefont {Ma}, \citenamefont {Hu}, \citenamefont {Li}, \citenamefont {Liu}, \citenamefont {Yao}, \citenamefont {Jia},\ and\ \citenamefont {Liu}}]{Ma_2021}%
  \BibitemOpen
  \bibfield  {author} {\bibinfo {author} {\bibfnamefont {H.-Y.}\ \bibnamefont {Ma}}, \bibinfo {author} {\bibfnamefont {M.}~\bibnamefont {Hu}}, \bibinfo {author} {\bibfnamefont {N.}~\bibnamefont {Li}}, \bibinfo {author} {\bibfnamefont {J.}~\bibnamefont {Liu}}, \bibinfo {author} {\bibfnamefont {W.}~\bibnamefont {Yao}}, \bibinfo {author} {\bibfnamefont {J.-F.}\ \bibnamefont {Jia}},\ and\ \bibinfo {author} {\bibfnamefont {J.}~\bibnamefont {Liu}},\ }\bibfield  {title} {\bibinfo {title} {Multifunctional antiferromagnetic materials with giant piezomagnetism and noncollinear spin current},\ }\href {https://doi.org/10.1038/s41467-021-23127-7} {\bibfield  {journal} {\bibinfo  {journal} {Nat. Commun.}\ }\textbf {\bibinfo {volume} {12}},\ \bibinfo {pages} {2846} (\bibinfo {year} {2021})}\BibitemShut {NoStop}%
\bibitem [{\citenamefont {\ifmmode~\check{S}\else \v{S}\fi{}mejkal}\ \emph {et~al.}(2022{\natexlab{a}})\citenamefont {\ifmmode~\check{S}\else \v{S}\fi{}mejkal}, \citenamefont {Sinova},\ and\ \citenamefont {Jungwirth}}]{PhysRevX.12.040501}%
  \BibitemOpen
  \bibfield  {author} {\bibinfo {author} {\bibfnamefont {L.}~\bibnamefont {\ifmmode~\check{S}\else \v{S}\fi{}mejkal}}, \bibinfo {author} {\bibfnamefont {J.}~\bibnamefont {Sinova}},\ and\ \bibinfo {author} {\bibfnamefont {T.}~\bibnamefont {Jungwirth}},\ }\bibfield  {title} {\bibinfo {title} {{Emerging Research Landscape of Altermagnetism}},\ }\href {https://doi.org/10.1103/PhysRevX.12.040501} {\bibfield  {journal} {\bibinfo  {journal} {Phys. Rev. X}\ }\textbf {\bibinfo {volume} {12}},\ \bibinfo {pages} {040501} (\bibinfo {year} {2022}{\natexlab{a}})}\BibitemShut {NoStop}%
\bibitem [{\citenamefont {\ifmmode~\check{S}\else \v{S}\fi{}mejkal}\ \emph {et~al.}(2022{\natexlab{b}})\citenamefont {\ifmmode~\check{S}\else \v{S}\fi{}mejkal}, \citenamefont {Hellenes}, \citenamefont {Gonz\'alez-Hern\'andez}, \citenamefont {Sinova},\ and\ \citenamefont {Jungwirth}}]{PhysRevX.12.011028}%
  \BibitemOpen
  \bibfield  {author} {\bibinfo {author} {\bibfnamefont {L.}~\bibnamefont {\ifmmode~\check{S}\else \v{S}\fi{}mejkal}}, \bibinfo {author} {\bibfnamefont {A.~B.}\ \bibnamefont {Hellenes}}, \bibinfo {author} {\bibfnamefont {R.}~\bibnamefont {Gonz\'alez-Hern\'andez}}, \bibinfo {author} {\bibfnamefont {J.}~\bibnamefont {Sinova}},\ and\ \bibinfo {author} {\bibfnamefont {T.}~\bibnamefont {Jungwirth}},\ }\bibfield  {title} {\bibinfo {title} {{Giant and Tunneling Magnetoresistance in Unconventional Collinear Antiferromagnets with Nonrelativistic Spin-Momentum Coupling}},\ }\href {https://doi.org/10.1103/PhysRevX.12.011028} {\bibfield  {journal} {\bibinfo  {journal} {Phys. Rev. X}\ }\textbf {\bibinfo {volume} {12}},\ \bibinfo {pages} {011028} (\bibinfo {year} {2022}{\natexlab{b}})}\BibitemShut {NoStop}%
\bibitem [{\citenamefont {\ifmmode~\check{S}\else \v{S}\fi{}mejkal}\ \emph {et~al.}(2022{\natexlab{c}})\citenamefont {\ifmmode~\check{S}\else \v{S}\fi{}mejkal}, \citenamefont {Sinova},\ and\ \citenamefont {Jungwirth}}]{PhysRevX.12.031042}%
  \BibitemOpen
  \bibfield  {author} {\bibinfo {author} {\bibfnamefont {L.}~\bibnamefont {\ifmmode~\check{S}\else \v{S}\fi{}mejkal}}, \bibinfo {author} {\bibfnamefont {J.}~\bibnamefont {Sinova}},\ and\ \bibinfo {author} {\bibfnamefont {T.}~\bibnamefont {Jungwirth}},\ }\bibfield  {title} {\bibinfo {title} {{Beyond Conventional Ferromagnetism and Antiferromagnetism: A Phase with Nonrelativistic Spin and Crystal Rotation Symmetry}},\ }\href {https://doi.org/10.1103/PhysRevX.12.031042} {\bibfield  {journal} {\bibinfo  {journal} {Phys. Rev. X}\ }\textbf {\bibinfo {volume} {12}},\ \bibinfo {pages} {031042} (\bibinfo {year} {2022}{\natexlab{c}})}\BibitemShut {NoStop}%
\bibitem [{\citenamefont {\ifmmode~\check{S}\else \v{S}\fi{}mejkal}\ \emph {et~al.}(2023)\citenamefont {\ifmmode~\check{S}\else \v{S}\fi{}mejkal}, \citenamefont {Marmodoro}, \citenamefont {Ahn}, \citenamefont {Gonz\'alez-Hern\'andez}, \citenamefont {Turek}, \citenamefont {Mankovsky}, \citenamefont {Ebert}, \citenamefont {D'Souza}, \citenamefont {\ifmmode~\check{S}\else \v{S}\fi{}ipr}, \citenamefont {Sinova},\ and\ \citenamefont {Jungwirth}}]{PhysRevLett.131.256703}%
  \BibitemOpen
  \bibfield  {author} {\bibinfo {author} {\bibfnamefont {L.}~\bibnamefont {\ifmmode~\check{S}\else \v{S}\fi{}mejkal}}, \bibinfo {author} {\bibfnamefont {A.}~\bibnamefont {Marmodoro}}, \bibinfo {author} {\bibfnamefont {K.-H.}\ \bibnamefont {Ahn}}, \bibinfo {author} {\bibfnamefont {R.}~\bibnamefont {Gonz\'alez-Hern\'andez}}, \bibinfo {author} {\bibfnamefont {I.}~\bibnamefont {Turek}}, \bibinfo {author} {\bibfnamefont {S.}~\bibnamefont {Mankovsky}}, \bibinfo {author} {\bibfnamefont {H.}~\bibnamefont {Ebert}}, \bibinfo {author} {\bibfnamefont {S.~W.}\ \bibnamefont {D'Souza}}, \bibinfo {author} {\bibfnamefont {O.}~\bibnamefont {\ifmmode~\check{S}\else \v{S}\fi{}ipr}}, \bibinfo {author} {\bibfnamefont {J.}~\bibnamefont {Sinova}},\ and\ \bibinfo {author} {\bibfnamefont {T.}~\bibnamefont {Jungwirth}},\ }\bibfield  {title} {\bibinfo {title} {{Chiral Magnons in Altermagnetic ${\mathrm{RuO}}_{2}$}},\ }\href {https://doi.org/10.1103/PhysRevLett.131.256703} {\bibfield  {journal} {\bibinfo  {journal} {Phys. Rev. Lett.}\
  }\textbf {\bibinfo {volume} {131}},\ \bibinfo {pages} {256703} (\bibinfo {year} {2023})}\BibitemShut {NoStop}%
\bibitem [{\citenamefont {Yuan}\ and\ \citenamefont {Zunger}(2023)}]{Zunger-review}%
  \BibitemOpen
  \bibfield  {author} {\bibinfo {author} {\bibfnamefont {L.-D.}\ \bibnamefont {Yuan}}\ and\ \bibinfo {author} {\bibfnamefont {A.}~\bibnamefont {Zunger}},\ }\bibfield  {title} {\bibinfo {title} {Degeneracy removal of spin bands in collinear antiferromagnets with non-interconvertible spin-structure motif pair},\ }\href {https://doi.org/10.1002/adma.202211966} {\bibfield  {journal} {\bibinfo  {journal} {Adv. Mater.}\ }\textbf {\bibinfo {volume} {35}},\ \bibinfo {pages} {2211966} (\bibinfo {year} {2023})}\BibitemShut {NoStop}%
\bibitem [{\citenamefont {Yuan}\ \emph {et~al.}(2023)\citenamefont {Yuan}, \citenamefont {Zhang}, \citenamefont {Acosta},\ and\ \citenamefont {Zunger}}]{yuan2023uncovering}%
  \BibitemOpen
  \bibfield  {author} {\bibinfo {author} {\bibfnamefont {L.-D.}\ \bibnamefont {Yuan}}, \bibinfo {author} {\bibfnamefont {X.}~\bibnamefont {Zhang}}, \bibinfo {author} {\bibfnamefont {C.~M.}\ \bibnamefont {Acosta}},\ and\ \bibinfo {author} {\bibfnamefont {A.}~\bibnamefont {Zunger}},\ }\bibfield  {title} {\bibinfo {title} {{Uncovering spin-orbit coupling-independent hidden spin polarization of energy bands in antiferromagnets}},\ }\href {https://doi.org/10.1038/s41467-023-40877-8} {\bibfield  {journal} {\bibinfo  {journal} {Nat. Commun.}\ }\textbf {\bibinfo {volume} {14}},\ \bibinfo {pages} {5301} (\bibinfo {year} {2023})}\BibitemShut {NoStop}%
\bibitem [{\citenamefont {{Yan}}\ \emph {et~al.}(2024)\citenamefont {{Yan}}, \citenamefont {{Zhou}}, \citenamefont {{Qin}},\ and\ \citenamefont {{Liu}}}]{2024ApPhL.124c0503Y}%
  \BibitemOpen
  \bibfield  {author} {\bibinfo {author} {\bibfnamefont {H.}~\bibnamefont {{Yan}}}, \bibinfo {author} {\bibfnamefont {X.}~\bibnamefont {{Zhou}}}, \bibinfo {author} {\bibfnamefont {P.}~\bibnamefont {{Qin}}},\ and\ \bibinfo {author} {\bibfnamefont {Z.}~\bibnamefont {{Liu}}},\ }\bibfield  {title} {\bibinfo {title} {{Review on spin-split antiferromagnetic spintronics}},\ }\href {https://doi.org/10.1063/5.0184580} {\bibfield  {journal} {\bibinfo  {journal} {Appl. Phys. Lett.}\ }\textbf {\bibinfo {volume} {124}},\ \bibinfo {eid} {030503} (\bibinfo {year} {2024})}\BibitemShut {NoStop}%
\bibitem [{\citenamefont {Bhowal}\ and\ \citenamefont {Spaldin}(2024)}]{PhysRevX.14.011019}%
  \BibitemOpen
  \bibfield  {author} {\bibinfo {author} {\bibfnamefont {S.}~\bibnamefont {Bhowal}}\ and\ \bibinfo {author} {\bibfnamefont {N.~A.}\ \bibnamefont {Spaldin}},\ }\bibfield  {title} {\bibinfo {title} {{Ferroically Ordered Magnetic Octupoles in $d$-Wave Altermagnets}},\ }\href {https://doi.org/10.1103/PhysRevX.14.011019} {\bibfield  {journal} {\bibinfo  {journal} {Phys. Rev. X}\ }\textbf {\bibinfo {volume} {14}},\ \bibinfo {pages} {011019} (\bibinfo {year} {2024})}\BibitemShut {NoStop}%
\bibitem [{\citenamefont {Sødequist}\ and\ \citenamefont {Olsen}(2024)}]{S_dequist_2024}%
  \BibitemOpen
  \bibfield  {author} {\bibinfo {author} {\bibfnamefont {J.}~\bibnamefont {Sødequist}}\ and\ \bibinfo {author} {\bibfnamefont {T.}~\bibnamefont {Olsen}},\ }\bibfield  {title} {\bibinfo {title} {{Two-dimensional altermagnets from high throughput computational screening: Symmetry requirements, chiral magnons, and spin-orbit effects}},\ }\href {https://doi.org/10.1063/5.0198285} {\bibfield  {journal} {\bibinfo  {journal} {Appl. Phys. Lett.}\ }\textbf {\bibinfo {volume} {124}},\ \bibinfo {pages} {182409} (\bibinfo {year} {2024})}\BibitemShut {NoStop}%
\bibitem [{\citenamefont {Cheong}\ and\ \citenamefont {Huang}(2024)}]{Cheong_2024}%
  \BibitemOpen
  \bibfield  {author} {\bibinfo {author} {\bibfnamefont {S.-W.}\ \bibnamefont {Cheong}}\ and\ \bibinfo {author} {\bibfnamefont {F.-T.}\ \bibnamefont {Huang}},\ }\bibfield  {title} {\bibinfo {title} {Altermagnetism with non-collinear spins},\ }\href {https://doi.org/10.1038/s41535-024-00626-6} {\bibfield  {journal} {\bibinfo  {journal} {npj Quantum Mater.}\ }\textbf {\bibinfo {volume} {9}},\ \bibinfo {pages} {13} (\bibinfo {year} {2024})}\BibitemShut {NoStop}%
\bibitem [{\citenamefont {Kimel}\ \emph {et~al.}(2024)\citenamefont {Kimel}, \citenamefont {Rasing},\ and\ \citenamefont {Ivanov}}]{KIMEL2024172039}%
  \BibitemOpen
  \bibfield  {author} {\bibinfo {author} {\bibfnamefont {A.}~\bibnamefont {Kimel}}, \bibinfo {author} {\bibfnamefont {T.}~\bibnamefont {Rasing}},\ and\ \bibinfo {author} {\bibfnamefont {B.}~\bibnamefont {Ivanov}},\ }\bibfield  {title} {\bibinfo {title} {{Optical read-out and control of antiferromagnetic Néel vector in altermagnets and beyond}},\ }\href {https://doi.org/10.1016/j.jmmm.2024.172039} {\bibfield  {journal} {\bibinfo  {journal} {J. Magn. Magn. Mater.}\ }\textbf {\bibinfo {volume} {598}},\ \bibinfo {pages} {172039} (\bibinfo {year} {2024})}\BibitemShut {NoStop}%
\bibitem [{\citenamefont {Bai}\ \emph {et~al.}(2024)\citenamefont {Bai}, \citenamefont {Feng}, \citenamefont {Liu}, \citenamefont {Šmejkal}, \citenamefont {Mokrousov},\ and\ \citenamefont {Yao}}]{https://doi.org/10.1002/adfm.202409327}%
  \BibitemOpen
  \bibfield  {author} {\bibinfo {author} {\bibfnamefont {L.}~\bibnamefont {Bai}}, \bibinfo {author} {\bibfnamefont {W.}~\bibnamefont {Feng}}, \bibinfo {author} {\bibfnamefont {S.}~\bibnamefont {Liu}}, \bibinfo {author} {\bibfnamefont {L.}~\bibnamefont {Šmejkal}}, \bibinfo {author} {\bibfnamefont {Y.}~\bibnamefont {Mokrousov}},\ and\ \bibinfo {author} {\bibfnamefont {Y.}~\bibnamefont {Yao}},\ }\bibfield  {title} {\bibinfo {title} {{Altermagnetism: Exploring New Frontiers in Magnetism and Spintronics}},\ }\href {https://doi.org/https://doi.org/10.1002/adfm.202409327} {\bibfield  {journal} {\bibinfo  {journal} {Adv. Funct. Mater.}\ ,\ \bibinfo {pages} {2409327}} (\bibinfo {year} {2024})}\BibitemShut {NoStop}%
\bibitem [{\citenamefont {Jungwirth}\ \emph {et~al.}(2024{\natexlab{a}})\citenamefont {Jungwirth}, \citenamefont {Fernandes}, \citenamefont {Sinova},\ and\ \citenamefont {Smejkal}}]{jungwirth2024}%
  \BibitemOpen
  \bibfield  {author} {\bibinfo {author} {\bibfnamefont {T.}~\bibnamefont {Jungwirth}}, \bibinfo {author} {\bibfnamefont {R.~M.}\ \bibnamefont {Fernandes}}, \bibinfo {author} {\bibfnamefont {J.}~\bibnamefont {Sinova}},\ and\ \bibinfo {author} {\bibfnamefont {L.}~\bibnamefont {Smejkal}},\ }\href {https://arxiv.org/abs/2409.10034} {\bibinfo {title} {{Altermagnets and beyond: Nodal magnetically-ordered phases}}} (\bibinfo {year} {2024}{\natexlab{a}}),\ \Eprint {https://arxiv.org/abs/2409.10034} {arXiv:2409.10034 [cond-mat.mtrl-sci]} \BibitemShut {NoStop}%
\bibitem [{\citenamefont {Jungwirth}\ \emph {et~al.}(2024{\natexlab{b}})\citenamefont {Jungwirth}, \citenamefont {Fernandes}, \citenamefont {Fradkin}, \citenamefont {MacDonald}, \citenamefont {Sinova},\ and\ \citenamefont {Smejkal}}]{jungwirth2024supefluid3healtermagnets}%
  \BibitemOpen
  \bibfield  {author} {\bibinfo {author} {\bibfnamefont {T.}~\bibnamefont {Jungwirth}}, \bibinfo {author} {\bibfnamefont {R.~M.}\ \bibnamefont {Fernandes}}, \bibinfo {author} {\bibfnamefont {E.}~\bibnamefont {Fradkin}}, \bibinfo {author} {\bibfnamefont {A.~H.}\ \bibnamefont {MacDonald}}, \bibinfo {author} {\bibfnamefont {J.}~\bibnamefont {Sinova}},\ and\ \bibinfo {author} {\bibfnamefont {L.}~\bibnamefont {Smejkal}},\ }\href {https://arxiv.org/abs/2411.00717} {\bibinfo {title} {From supefluid 3he to altermagnets}} (\bibinfo {year} {2024}{\natexlab{b}}),\ \Eprint {https://arxiv.org/abs/2411.00717} {arXiv:2411.00717 [cond-mat.mtrl-sci]} \BibitemShut {NoStop}%
\bibitem [{\citenamefont {Krempask{\'y}}\ \emph {et~al.}(2024)\citenamefont {Krempask{\'y}}, \citenamefont {{\v{S}}mejkal}, \citenamefont {D'Souza}, \citenamefont {Hajlaoui}, \citenamefont {Springholz}, \citenamefont {Uhl{\'i}{\v{r}}ov{\'a}}, \citenamefont {Alarab}, \citenamefont {Constantinou}, \citenamefont {Strocov}, \citenamefont {Usanov}, \citenamefont {Pudelko}, \citenamefont {Gonz{\'a}lez-Hern{\'a}ndez}, \citenamefont {Birk~Hellenes}, \citenamefont {Jansa}, \citenamefont {Reichlov{\'a}}, \citenamefont {{\v{S}}ob{\'a}{\v{n}}}, \citenamefont {Gonzalez~Betancourt}, \citenamefont {Wadley}, \citenamefont {Sinova}, \citenamefont {Kriegner}, \citenamefont {Min{\'a}r}, \citenamefont {Dil},\ and\ \citenamefont {Jungwirth}}]{Krempasky2024}%
  \BibitemOpen
  \bibfield  {author} {\bibinfo {author} {\bibfnamefont {J.}~\bibnamefont {Krempask{\'y}}}, \bibinfo {author} {\bibfnamefont {L.}~\bibnamefont {{\v{S}}mejkal}}, \bibinfo {author} {\bibfnamefont {S.~W.}\ \bibnamefont {D'Souza}}, \bibinfo {author} {\bibfnamefont {M.}~\bibnamefont {Hajlaoui}}, \bibinfo {author} {\bibfnamefont {G.}~\bibnamefont {Springholz}}, \bibinfo {author} {\bibfnamefont {K.}~\bibnamefont {Uhl{\'i}{\v{r}}ov{\'a}}}, \bibinfo {author} {\bibfnamefont {F.}~\bibnamefont {Alarab}}, \bibinfo {author} {\bibfnamefont {P.~C.}\ \bibnamefont {Constantinou}}, \bibinfo {author} {\bibfnamefont {V.}~\bibnamefont {Strocov}}, \bibinfo {author} {\bibfnamefont {D.}~\bibnamefont {Usanov}}, \bibinfo {author} {\bibfnamefont {W.~R.}\ \bibnamefont {Pudelko}}, \bibinfo {author} {\bibfnamefont {R.}~\bibnamefont {Gonz{\'a}lez-Hern{\'a}ndez}}, \bibinfo {author} {\bibfnamefont {A.}~\bibnamefont {Birk~Hellenes}}, \bibinfo {author} {\bibfnamefont {Z.}~\bibnamefont {Jansa}}, \bibinfo {author} {\bibfnamefont {H.}~\bibnamefont
  {Reichlov{\'a}}}, \bibinfo {author} {\bibfnamefont {Z.}~\bibnamefont {{\v{S}}ob{\'a}{\v{n}}}}, \bibinfo {author} {\bibfnamefont {R.~D.}\ \bibnamefont {Gonzalez~Betancourt}}, \bibinfo {author} {\bibfnamefont {P.}~\bibnamefont {Wadley}}, \bibinfo {author} {\bibfnamefont {J.}~\bibnamefont {Sinova}}, \bibinfo {author} {\bibfnamefont {D.}~\bibnamefont {Kriegner}}, \bibinfo {author} {\bibfnamefont {J.}~\bibnamefont {Min{\'a}r}}, \bibinfo {author} {\bibfnamefont {J.~H.}\ \bibnamefont {Dil}},\ and\ \bibinfo {author} {\bibfnamefont {T.}~\bibnamefont {Jungwirth}},\ }\bibfield  {title} {\bibinfo {title} {{Altermagnetic lifting of Kramers spin degeneracy}},\ }\href {https://doi.org/10.1038/s41586-023-06907-7} {\bibfield  {journal} {\bibinfo  {journal} {Nature}\ }\textbf {\bibinfo {volume} {626}},\ \bibinfo {pages} {517} (\bibinfo {year} {2024})}\BibitemShut {NoStop}%
\bibitem [{\citenamefont {Lee}\ \emph {et~al.}(2024)\citenamefont {Lee}, \citenamefont {Lee}, \citenamefont {Jung}, \citenamefont {Jung}, \citenamefont {Kim}, \citenamefont {Lee}, \citenamefont {Seok}, \citenamefont {Kim}, \citenamefont {Park}, \citenamefont {\ifmmode~\check{S}\else \v{S}\fi{}mejkal}, \citenamefont {Kang},\ and\ \citenamefont {Kim}}]{PhysRevLett.132.036702}%
  \BibitemOpen
  \bibfield  {author} {\bibinfo {author} {\bibfnamefont {S.}~\bibnamefont {Lee}}, \bibinfo {author} {\bibfnamefont {S.}~\bibnamefont {Lee}}, \bibinfo {author} {\bibfnamefont {S.}~\bibnamefont {Jung}}, \bibinfo {author} {\bibfnamefont {J.}~\bibnamefont {Jung}}, \bibinfo {author} {\bibfnamefont {D.}~\bibnamefont {Kim}}, \bibinfo {author} {\bibfnamefont {Y.}~\bibnamefont {Lee}}, \bibinfo {author} {\bibfnamefont {B.}~\bibnamefont {Seok}}, \bibinfo {author} {\bibfnamefont {J.}~\bibnamefont {Kim}}, \bibinfo {author} {\bibfnamefont {B.~G.}\ \bibnamefont {Park}}, \bibinfo {author} {\bibfnamefont {L.}~\bibnamefont {\ifmmode~\check{S}\else \v{S}\fi{}mejkal}}, \bibinfo {author} {\bibfnamefont {C.-J.}\ \bibnamefont {Kang}},\ and\ \bibinfo {author} {\bibfnamefont {C.}~\bibnamefont {Kim}},\ }\bibfield  {title} {\bibinfo {title} {{Broken Kramers Degeneracy in Altermagnetic MnTe}},\ }\href {https://doi.org/10.1103/PhysRevLett.132.036702} {\bibfield  {journal} {\bibinfo  {journal} {Phys. Rev. Lett.}\ }\textbf {\bibinfo
  {volume} {132}},\ \bibinfo {pages} {036702} (\bibinfo {year} {2024})}\BibitemShut {NoStop}%
\bibitem [{\citenamefont {Osumi}\ \emph {et~al.}(2024)\citenamefont {Osumi}, \citenamefont {Souma}, \citenamefont {Aoyama}, \citenamefont {Yamauchi}, \citenamefont {Honma}, \citenamefont {Nakayama}, \citenamefont {Takahashi}, \citenamefont {Ohgushi},\ and\ \citenamefont {Sato}}]{PhysRevB.109.115102}%
  \BibitemOpen
  \bibfield  {author} {\bibinfo {author} {\bibfnamefont {T.}~\bibnamefont {Osumi}}, \bibinfo {author} {\bibfnamefont {S.}~\bibnamefont {Souma}}, \bibinfo {author} {\bibfnamefont {T.}~\bibnamefont {Aoyama}}, \bibinfo {author} {\bibfnamefont {K.}~\bibnamefont {Yamauchi}}, \bibinfo {author} {\bibfnamefont {A.}~\bibnamefont {Honma}}, \bibinfo {author} {\bibfnamefont {K.}~\bibnamefont {Nakayama}}, \bibinfo {author} {\bibfnamefont {T.}~\bibnamefont {Takahashi}}, \bibinfo {author} {\bibfnamefont {K.}~\bibnamefont {Ohgushi}},\ and\ \bibinfo {author} {\bibfnamefont {T.}~\bibnamefont {Sato}},\ }\bibfield  {title} {\bibinfo {title} {{Observation of a giant band splitting in altermagnetic MnTe}},\ }\href {https://doi.org/10.1103/PhysRevB.109.115102} {\bibfield  {journal} {\bibinfo  {journal} {Phys. Rev. B}\ }\textbf {\bibinfo {volume} {109}},\ \bibinfo {pages} {115102} (\bibinfo {year} {2024})}\BibitemShut {NoStop}%
\bibitem [{\citenamefont {Fedchenko}\ \emph {et~al.}(2024)\citenamefont {Fedchenko}, \citenamefont {Minár}, \citenamefont {Akashdeep}, \citenamefont {D’Souza}, \citenamefont {Vasilyev}, \citenamefont {Tkach}, \citenamefont {Odenbreit}, \citenamefont {Nguyen}, \citenamefont {Kutnyakhov}, \citenamefont {Wind}, \citenamefont {Wenthaus}, \citenamefont {Scholz}, \citenamefont {Rossnagel}, \citenamefont {Hoesch}, \citenamefont {Aeschlimann}, \citenamefont {Stadtmüller}, \citenamefont {Kläui}, \citenamefont {Schönhense}, \citenamefont {Jungwirth}, \citenamefont {Hellenes}, \citenamefont {Jakob}, \citenamefont {Šmejkal}, \citenamefont {Sinova},\ and\ \citenamefont {Elmers}}]{doi:10.1126/sciadv.adj4883}%
  \BibitemOpen
  \bibfield  {author} {\bibinfo {author} {\bibfnamefont {O.}~\bibnamefont {Fedchenko}}, \bibinfo {author} {\bibfnamefont {J.}~\bibnamefont {Minár}}, \bibinfo {author} {\bibfnamefont {A.}~\bibnamefont {Akashdeep}}, \bibinfo {author} {\bibfnamefont {S.~W.}\ \bibnamefont {D’Souza}}, \bibinfo {author} {\bibfnamefont {D.}~\bibnamefont {Vasilyev}}, \bibinfo {author} {\bibfnamefont {O.}~\bibnamefont {Tkach}}, \bibinfo {author} {\bibfnamefont {L.}~\bibnamefont {Odenbreit}}, \bibinfo {author} {\bibfnamefont {Q.}~\bibnamefont {Nguyen}}, \bibinfo {author} {\bibfnamefont {D.}~\bibnamefont {Kutnyakhov}}, \bibinfo {author} {\bibfnamefont {N.}~\bibnamefont {Wind}}, \bibinfo {author} {\bibfnamefont {L.}~\bibnamefont {Wenthaus}}, \bibinfo {author} {\bibfnamefont {M.}~\bibnamefont {Scholz}}, \bibinfo {author} {\bibfnamefont {K.}~\bibnamefont {Rossnagel}}, \bibinfo {author} {\bibfnamefont {M.}~\bibnamefont {Hoesch}}, \bibinfo {author} {\bibfnamefont {M.}~\bibnamefont {Aeschlimann}}, \bibinfo {author} {\bibfnamefont
  {B.}~\bibnamefont {Stadtmüller}}, \bibinfo {author} {\bibfnamefont {M.}~\bibnamefont {Kläui}}, \bibinfo {author} {\bibfnamefont {G.}~\bibnamefont {Schönhense}}, \bibinfo {author} {\bibfnamefont {T.}~\bibnamefont {Jungwirth}}, \bibinfo {author} {\bibfnamefont {A.~B.}\ \bibnamefont {Hellenes}}, \bibinfo {author} {\bibfnamefont {G.}~\bibnamefont {Jakob}}, \bibinfo {author} {\bibfnamefont {L.}~\bibnamefont {Šmejkal}}, \bibinfo {author} {\bibfnamefont {J.}~\bibnamefont {Sinova}},\ and\ \bibinfo {author} {\bibfnamefont {H.-J.}\ \bibnamefont {Elmers}},\ }\bibfield  {title} {\bibinfo {title} {{Observation of time-reversal symmetry breaking in the band structure of altermagnetic RuO$_2$}},\ }\href {https://doi.org/10.1126/sciadv.adj4883} {\bibfield  {journal} {\bibinfo  {journal} {Sci. Adv.}\ }\textbf {\bibinfo {volume} {10}},\ \bibinfo {pages} {eadj4883} (\bibinfo {year} {2024})}\BibitemShut {NoStop}%
\bibitem [{\citenamefont {Reimers}\ \emph {et~al.}(2024)\citenamefont {Reimers}, \citenamefont {Odenbreit}, \citenamefont {Šmejkal}, \citenamefont {Strocov}, \citenamefont {Constantinou}, \citenamefont {Hellenes}, \citenamefont {Jaeschke~Ubiergo}, \citenamefont {Campos}, \citenamefont {Bharadwaj}, \citenamefont {Chakraborty}, \citenamefont {Denneulin}, \citenamefont {Shi}, \citenamefont {Dunin-Borkowski}, \citenamefont {Das}, \citenamefont {Kläui}, \citenamefont {Sinova},\ and\ \citenamefont {Jourdan}}]{Reimers_2024}%
  \BibitemOpen
  \bibfield  {author} {\bibinfo {author} {\bibfnamefont {S.}~\bibnamefont {Reimers}}, \bibinfo {author} {\bibfnamefont {L.}~\bibnamefont {Odenbreit}}, \bibinfo {author} {\bibfnamefont {L.}~\bibnamefont {Šmejkal}}, \bibinfo {author} {\bibfnamefont {V.~N.}\ \bibnamefont {Strocov}}, \bibinfo {author} {\bibfnamefont {P.}~\bibnamefont {Constantinou}}, \bibinfo {author} {\bibfnamefont {A.~B.}\ \bibnamefont {Hellenes}}, \bibinfo {author} {\bibfnamefont {R.}~\bibnamefont {Jaeschke~Ubiergo}}, \bibinfo {author} {\bibfnamefont {W.~H.}\ \bibnamefont {Campos}}, \bibinfo {author} {\bibfnamefont {V.~K.}\ \bibnamefont {Bharadwaj}}, \bibinfo {author} {\bibfnamefont {A.}~\bibnamefont {Chakraborty}}, \bibinfo {author} {\bibfnamefont {T.}~\bibnamefont {Denneulin}}, \bibinfo {author} {\bibfnamefont {W.}~\bibnamefont {Shi}}, \bibinfo {author} {\bibfnamefont {R.~E.}\ \bibnamefont {Dunin-Borkowski}}, \bibinfo {author} {\bibfnamefont {S.}~\bibnamefont {Das}}, \bibinfo {author} {\bibfnamefont {M.}~\bibnamefont {Kläui}}, \bibinfo
  {author} {\bibfnamefont {J.}~\bibnamefont {Sinova}},\ and\ \bibinfo {author} {\bibfnamefont {M.}~\bibnamefont {Jourdan}},\ }\bibfield  {title} {\bibinfo {title} {{Direct observation of altermagnetic band splitting in CrSb thin films}},\ }\href {https://doi.org/10.1038/s41467-024-46476-5} {\bibfield  {journal} {\bibinfo  {journal} {Nat. Commun.}\ }\textbf {\bibinfo {volume} {15}},\ \bibinfo {pages} {2116} (\bibinfo {year} {2024})}\BibitemShut {NoStop}%
\bibitem [{\citenamefont {Zeng}\ \emph {et~al.}(2024)\citenamefont {Zeng}, \citenamefont {Zhu}, \citenamefont {Zhu}, \citenamefont {Liu}, \citenamefont {Ma}, \citenamefont {Hao}, \citenamefont {Liu}, \citenamefont {Qu}, \citenamefont {Yang}, \citenamefont {Jiang}, \citenamefont {Yamagami}, \citenamefont {Arita}, \citenamefont {Zhang}, \citenamefont {Shao}, \citenamefont {Dai}, \citenamefont {Shimada}, \citenamefont {Liu}, \citenamefont {Ye}, \citenamefont {Huang}, \citenamefont {Liu},\ and\ \citenamefont {Liu}}]{https://doi.org/10.1002/advs.202406529}%
  \BibitemOpen
  \bibfield  {author} {\bibinfo {author} {\bibfnamefont {M.}~\bibnamefont {Zeng}}, \bibinfo {author} {\bibfnamefont {M.-Y.}\ \bibnamefont {Zhu}}, \bibinfo {author} {\bibfnamefont {Y.-P.}\ \bibnamefont {Zhu}}, \bibinfo {author} {\bibfnamefont {X.-R.}\ \bibnamefont {Liu}}, \bibinfo {author} {\bibfnamefont {X.-M.}\ \bibnamefont {Ma}}, \bibinfo {author} {\bibfnamefont {Y.-J.}\ \bibnamefont {Hao}}, \bibinfo {author} {\bibfnamefont {P.}~\bibnamefont {Liu}}, \bibinfo {author} {\bibfnamefont {G.}~\bibnamefont {Qu}}, \bibinfo {author} {\bibfnamefont {Y.}~\bibnamefont {Yang}}, \bibinfo {author} {\bibfnamefont {Z.}~\bibnamefont {Jiang}}, \bibinfo {author} {\bibfnamefont {K.}~\bibnamefont {Yamagami}}, \bibinfo {author} {\bibfnamefont {M.}~\bibnamefont {Arita}}, \bibinfo {author} {\bibfnamefont {X.}~\bibnamefont {Zhang}}, \bibinfo {author} {\bibfnamefont {T.-H.}\ \bibnamefont {Shao}}, \bibinfo {author} {\bibfnamefont {Y.}~\bibnamefont {Dai}}, \bibinfo {author} {\bibfnamefont {K.}~\bibnamefont {Shimada}}, \bibinfo {author}
  {\bibfnamefont {Z.}~\bibnamefont {Liu}}, \bibinfo {author} {\bibfnamefont {M.}~\bibnamefont {Ye}}, \bibinfo {author} {\bibfnamefont {Y.}~\bibnamefont {Huang}}, \bibinfo {author} {\bibfnamefont {Q.}~\bibnamefont {Liu}},\ and\ \bibinfo {author} {\bibfnamefont {C.}~\bibnamefont {Liu}},\ }\bibfield  {title} {\bibinfo {title} {{Observation of Spin Splitting in Room-Temperature Metallic Antiferromagnet CrSb}},\ }\href {https://doi.org/10.1002/advs.202406529} {\bibfield  {journal} {\bibinfo  {journal} {Adv. Sci.}\ ,\ \bibinfo {pages} {2406529}} (\bibinfo {year} {2024})}\BibitemShut {NoStop}%
\bibitem [{\citenamefont {Ding}\ \emph {et~al.}(2024)\citenamefont {Ding}, \citenamefont {Jiang}, \citenamefont {Chen}, \citenamefont {Tao}, \citenamefont {Liu}, \citenamefont {Li}, \citenamefont {Liu}, \citenamefont {Sun}, \citenamefont {Cheng}, \citenamefont {Liu}, \citenamefont {Yang}, \citenamefont {Zhang}, \citenamefont {Deng}, \citenamefont {Jing}, \citenamefont {Huang}, \citenamefont {Shi}, \citenamefont {Ye}, \citenamefont {Qiao}, \citenamefont {Wang}, \citenamefont {Guo}, \citenamefont {Feng},\ and\ \citenamefont {Shen}}]{ding2024largebandsplittinggwavetype}%
  \BibitemOpen
  \bibfield  {author} {\bibinfo {author} {\bibfnamefont {J.}~\bibnamefont {Ding}}, \bibinfo {author} {\bibfnamefont {Z.}~\bibnamefont {Jiang}}, \bibinfo {author} {\bibfnamefont {X.}~\bibnamefont {Chen}}, \bibinfo {author} {\bibfnamefont {Z.}~\bibnamefont {Tao}}, \bibinfo {author} {\bibfnamefont {Z.}~\bibnamefont {Liu}}, \bibinfo {author} {\bibfnamefont {T.}~\bibnamefont {Li}}, \bibinfo {author} {\bibfnamefont {J.}~\bibnamefont {Liu}}, \bibinfo {author} {\bibfnamefont {J.}~\bibnamefont {Sun}}, \bibinfo {author} {\bibfnamefont {J.}~\bibnamefont {Cheng}}, \bibinfo {author} {\bibfnamefont {J.}~\bibnamefont {Liu}}, \bibinfo {author} {\bibfnamefont {Y.}~\bibnamefont {Yang}}, \bibinfo {author} {\bibfnamefont {R.}~\bibnamefont {Zhang}}, \bibinfo {author} {\bibfnamefont {L.}~\bibnamefont {Deng}}, \bibinfo {author} {\bibfnamefont {W.}~\bibnamefont {Jing}}, \bibinfo {author} {\bibfnamefont {Y.}~\bibnamefont {Huang}}, \bibinfo {author} {\bibfnamefont {Y.}~\bibnamefont {Shi}}, \bibinfo {author} {\bibfnamefont
  {M.}~\bibnamefont {Ye}}, \bibinfo {author} {\bibfnamefont {S.}~\bibnamefont {Qiao}}, \bibinfo {author} {\bibfnamefont {Y.}~\bibnamefont {Wang}}, \bibinfo {author} {\bibfnamefont {Y.}~\bibnamefont {Guo}}, \bibinfo {author} {\bibfnamefont {D.}~\bibnamefont {Feng}},\ and\ \bibinfo {author} {\bibfnamefont {D.}~\bibnamefont {Shen}},\ }\bibfield  {title} {\bibinfo {title} {{Large Band Splitting in $g$-Wave Altermagnet CrSb}},\ }\href {https://doi.org/10.1103/PhysRevLett.133.206401} {\bibfield  {journal} {\bibinfo  {journal} {Phys. Rev. Lett.}\ }\textbf {\bibinfo {volume} {133}},\ \bibinfo {pages} {206401} (\bibinfo {year} {2024})}\BibitemShut {NoStop}%
\bibitem [{\citenamefont {Yuan}\ \emph {et~al.}(2024)\citenamefont {Yuan}, \citenamefont {Georgescu},\ and\ \citenamefont {Rondinelli}}]{PhysRevLett.133.216701}%
  \BibitemOpen
  \bibfield  {author} {\bibinfo {author} {\bibfnamefont {L.-D.}\ \bibnamefont {Yuan}}, \bibinfo {author} {\bibfnamefont {A.~B.}\ \bibnamefont {Georgescu}},\ and\ \bibinfo {author} {\bibfnamefont {J.~M.}\ \bibnamefont {Rondinelli}},\ }\bibfield  {title} {\bibinfo {title} {{Nonrelativistic Spin Splitting at the Brillouin Zone Center in Compensated Magnets}},\ }\href {https://doi.org/10.1103/PhysRevLett.133.216701} {\bibfield  {journal} {\bibinfo  {journal} {Phys. Rev. Lett.}\ }\textbf {\bibinfo {volume} {133}},\ \bibinfo {pages} {216701} (\bibinfo {year} {2024})}\BibitemShut {NoStop}%
\bibitem [{\citenamefont {Gong}\ \emph {et~al.}(2018)\citenamefont {Gong}, \citenamefont {Gong}, \citenamefont {Sun}, \citenamefont {Tong}, \citenamefont {Duan}, \citenamefont {Chu},\ and\ \citenamefont {Zhang}}]{doi:10.1073/pnas.1715465115}%
  \BibitemOpen
  \bibfield  {author} {\bibinfo {author} {\bibfnamefont {S.-J.}\ \bibnamefont {Gong}}, \bibinfo {author} {\bibfnamefont {C.}~\bibnamefont {Gong}}, \bibinfo {author} {\bibfnamefont {Y.-Y.}\ \bibnamefont {Sun}}, \bibinfo {author} {\bibfnamefont {W.-Y.}\ \bibnamefont {Tong}}, \bibinfo {author} {\bibfnamefont {C.-G.}\ \bibnamefont {Duan}}, \bibinfo {author} {\bibfnamefont {J.-H.}\ \bibnamefont {Chu}},\ and\ \bibinfo {author} {\bibfnamefont {X.}~\bibnamefont {Zhang}},\ }\bibfield  {title} {\bibinfo {title} {{Electrically induced 2D half-metallic antiferromagnets and spin field effect transistors}},\ }\href {https://doi.org/10.1073/pnas.1715465115} {\bibfield  {journal} {\bibinfo  {journal} {Proc. Natl. Acad. Sci. U.S.A.}\ }\textbf {\bibinfo {volume} {115}},\ \bibinfo {pages} {8511} (\bibinfo {year} {2018})}\BibitemShut {NoStop}%
\bibitem [{\citenamefont {van Leuken}\ and\ \citenamefont {de~Groot}(1995)}]{PhysRevLett.74.1171}%
  \BibitemOpen
  \bibfield  {author} {\bibinfo {author} {\bibfnamefont {H.}~\bibnamefont {van Leuken}}\ and\ \bibinfo {author} {\bibfnamefont {R.~A.}\ \bibnamefont {de~Groot}},\ }\bibfield  {title} {\bibinfo {title} {Half-metallic antiferromagnets},\ }\href {https://doi.org/10.1103/PhysRevLett.74.1171} {\bibfield  {journal} {\bibinfo  {journal} {Phys. Rev. Lett.}\ }\textbf {\bibinfo {volume} {74}},\ \bibinfo {pages} {1171} (\bibinfo {year} {1995})}\BibitemShut {NoStop}%
\bibitem [{\citenamefont {Pekar}\ and\ \citenamefont {Rashba}(1965)}]{osti_4642614}%
  \BibitemOpen
  \bibfield  {author} {\bibinfo {author} {\bibfnamefont {S.~I.}\ \bibnamefont {Pekar}}\ and\ \bibinfo {author} {\bibfnamefont {E.~I.}\ \bibnamefont {Rashba}},\ }\bibfield  {title} {\bibinfo {title} {Combined resonance in crystals in inhomogeneous magnetic fields},\ }\href {https://www.jetp.ras.ru/cgi-bin/e/index/r/47/5/p1927?a=list} {\bibfield  {journal} {\bibinfo  {journal} {J. Exp. Theor. Phys.}\ }\textbf {\bibinfo {volume} {20}},\ \bibinfo {pages} {1295} (\bibinfo {year} {1965})}\BibitemShut {NoStop}%
\bibitem [{\citenamefont {Kriegner}\ \emph {et~al.}(2016)\citenamefont {Kriegner}, \citenamefont {Výborný}, \citenamefont {Olejník}, \citenamefont {Reichlová}, \citenamefont {Novák}, \citenamefont {Marti}, \citenamefont {Gazquez}, \citenamefont {Saidl}, \citenamefont {Němec}, \citenamefont {Volobuev}, \citenamefont {Springholz}, \citenamefont {Holý},\ and\ \citenamefont {Jungwirth}}]{Kriegner_2016}%
  \BibitemOpen
  \bibfield  {author} {\bibinfo {author} {\bibfnamefont {D.}~\bibnamefont {Kriegner}}, \bibinfo {author} {\bibfnamefont {K.}~\bibnamefont {Výborný}}, \bibinfo {author} {\bibfnamefont {K.}~\bibnamefont {Olejník}}, \bibinfo {author} {\bibfnamefont {H.}~\bibnamefont {Reichlová}}, \bibinfo {author} {\bibfnamefont {V.}~\bibnamefont {Novák}}, \bibinfo {author} {\bibfnamefont {X.}~\bibnamefont {Marti}}, \bibinfo {author} {\bibfnamefont {J.}~\bibnamefont {Gazquez}}, \bibinfo {author} {\bibfnamefont {V.}~\bibnamefont {Saidl}}, \bibinfo {author} {\bibfnamefont {P.}~\bibnamefont {Němec}}, \bibinfo {author} {\bibfnamefont {V.~V.}\ \bibnamefont {Volobuev}}, \bibinfo {author} {\bibfnamefont {G.}~\bibnamefont {Springholz}}, \bibinfo {author} {\bibfnamefont {V.}~\bibnamefont {Holý}},\ and\ \bibinfo {author} {\bibfnamefont {T.}~\bibnamefont {Jungwirth}},\ }\bibfield  {title} {\bibinfo {title} {{Multiple-stable anisotropic magnetoresistance memory in antiferromagnetic MnTe}},\ }\href
  {https://doi.org/10.1038/ncomms11623} {\bibfield  {journal} {\bibinfo  {journal} {Nat. Commun.}\ }\textbf {\bibinfo {volume} {7}},\ \bibinfo {pages} {11623} (\bibinfo {year} {2016})}\BibitemShut {NoStop}%
\bibitem [{\citenamefont {Bossini}\ \emph {et~al.}(2020)\citenamefont {Bossini}, \citenamefont {Terschanski}, \citenamefont {Mertens}, \citenamefont {Springholz}, \citenamefont {Bonanni}, \citenamefont {Uhrig},\ and\ \citenamefont {Cinchetti}}]{bossini2020exchange}%
  \BibitemOpen
  \bibfield  {author} {\bibinfo {author} {\bibfnamefont {D.}~\bibnamefont {Bossini}}, \bibinfo {author} {\bibfnamefont {M.}~\bibnamefont {Terschanski}}, \bibinfo {author} {\bibfnamefont {F.}~\bibnamefont {Mertens}}, \bibinfo {author} {\bibfnamefont {G.}~\bibnamefont {Springholz}}, \bibinfo {author} {\bibfnamefont {A.}~\bibnamefont {Bonanni}}, \bibinfo {author} {\bibfnamefont {G.~S.}\ \bibnamefont {Uhrig}},\ and\ \bibinfo {author} {\bibfnamefont {M.}~\bibnamefont {Cinchetti}},\ }\bibfield  {title} {\bibinfo {title} {{Exchange-mediated magnetic blue-shift of the band-gap energy in the antiferromagnetic semiconductor MnTe}},\ }\href {https://doi.org/10.1088/1367-2630/aba0e7} {\bibfield  {journal} {\bibinfo  {journal} {New J. Phys.}\ }\textbf {\bibinfo {volume} {22}},\ \bibinfo {pages} {083029} (\bibinfo {year} {2020})}\BibitemShut {NoStop}%
\bibitem [{\citenamefont {Devaraj}\ \emph {et~al.}(2024)\citenamefont {Devaraj}, \citenamefont {Bose},\ and\ \citenamefont {Narayan}}]{PhysRevMaterials.8.104407}%
  \BibitemOpen
  \bibfield  {author} {\bibinfo {author} {\bibfnamefont {N.}~\bibnamefont {Devaraj}}, \bibinfo {author} {\bibfnamefont {A.}~\bibnamefont {Bose}},\ and\ \bibinfo {author} {\bibfnamefont {A.}~\bibnamefont {Narayan}},\ }\bibfield  {title} {\bibinfo {title} {{Interplay of altermagnetism and pressure in hexagonal and orthorhombic MnTe}},\ }\href {https://doi.org/10.1103/PhysRevMaterials.8.104407} {\bibfield  {journal} {\bibinfo  {journal} {Phys. Rev. Mater.}\ }\textbf {\bibinfo {volume} {8}},\ \bibinfo {pages} {104407} (\bibinfo {year} {2024})}\BibitemShut {NoStop}%
\bibitem [{\citenamefont {Hariki}\ \emph {et~al.}(2024)\citenamefont {Hariki}, \citenamefont {Dal~Din}, \citenamefont {Amin}, \citenamefont {Yamaguchi}, \citenamefont {Badura}, \citenamefont {Kriegner}, \citenamefont {Edmonds}, \citenamefont {Campion}, \citenamefont {Wadley}, \citenamefont {Backes}, \citenamefont {Veiga}, \citenamefont {Dhesi}, \citenamefont {Springholz}, \citenamefont {\ifmmode~\check{S}\else \v{S}\fi{}mejkal}, \citenamefont {V\'yborn\'y}, \citenamefont {Jungwirth},\ and\ \citenamefont {Kune\ifmmode~\check{s}\else \v{s}\fi{}}}]{PhysRevLett.132.176701}%
  \BibitemOpen
  \bibfield  {author} {\bibinfo {author} {\bibfnamefont {A.}~\bibnamefont {Hariki}}, \bibinfo {author} {\bibfnamefont {A.}~\bibnamefont {Dal~Din}}, \bibinfo {author} {\bibfnamefont {O.~J.}\ \bibnamefont {Amin}}, \bibinfo {author} {\bibfnamefont {T.}~\bibnamefont {Yamaguchi}}, \bibinfo {author} {\bibfnamefont {A.}~\bibnamefont {Badura}}, \bibinfo {author} {\bibfnamefont {D.}~\bibnamefont {Kriegner}}, \bibinfo {author} {\bibfnamefont {K.~W.}\ \bibnamefont {Edmonds}}, \bibinfo {author} {\bibfnamefont {R.~P.}\ \bibnamefont {Campion}}, \bibinfo {author} {\bibfnamefont {P.}~\bibnamefont {Wadley}}, \bibinfo {author} {\bibfnamefont {D.}~\bibnamefont {Backes}}, \bibinfo {author} {\bibfnamefont {L.~S.~I.}\ \bibnamefont {Veiga}}, \bibinfo {author} {\bibfnamefont {S.~S.}\ \bibnamefont {Dhesi}}, \bibinfo {author} {\bibfnamefont {G.}~\bibnamefont {Springholz}}, \bibinfo {author} {\bibfnamefont {L.}~\bibnamefont {\ifmmode~\check{S}\else \v{S}\fi{}mejkal}}, \bibinfo {author} {\bibfnamefont {K.}~\bibnamefont {V\'yborn\'y}},
  \bibinfo {author} {\bibfnamefont {T.}~\bibnamefont {Jungwirth}},\ and\ \bibinfo {author} {\bibfnamefont {J.}~\bibnamefont {Kune\ifmmode~\check{s}\else \v{s}\fi{}}},\ }\bibfield  {title} {\bibinfo {title} {{X-Ray Magnetic Circular Dichroism in Altermagnetic $\ensuremath{\alpha}$-MnTe}},\ }\href {https://doi.org/10.1103/PhysRevLett.132.176701} {\bibfield  {journal} {\bibinfo  {journal} {Phys. Rev. Lett.}\ }\textbf {\bibinfo {volume} {132}},\ \bibinfo {pages} {176701} (\bibinfo {year} {2024})}\BibitemShut {NoStop}%
\bibitem [{\citenamefont {Chilcote}\ \emph {et~al.}(2024)\citenamefont {Chilcote}, \citenamefont {Mazza}, \citenamefont {Lu}, \citenamefont {Gray}, \citenamefont {Tian}, \citenamefont {Deng}, \citenamefont {Moseley}, \citenamefont {Chen}, \citenamefont {Lapano}, \citenamefont {Gardner}, \citenamefont {Eres}, \citenamefont {Ward}, \citenamefont {Feng}, \citenamefont {Cao}, \citenamefont {Lauter}, \citenamefont {McGuire}, \citenamefont {Hermann}, \citenamefont {Parker}, \citenamefont {Han}, \citenamefont {Kayani}, \citenamefont {Rimal}, \citenamefont {Wu}, \citenamefont {Charlton}, \citenamefont {Moore},\ and\ \citenamefont {Brahlek}}]{Chilcote_2024}%
  \BibitemOpen
  \bibfield  {author} {\bibinfo {author} {\bibfnamefont {M.}~\bibnamefont {Chilcote}}, \bibinfo {author} {\bibfnamefont {A.~R.}\ \bibnamefont {Mazza}}, \bibinfo {author} {\bibfnamefont {Q.}~\bibnamefont {Lu}}, \bibinfo {author} {\bibfnamefont {I.}~\bibnamefont {Gray}}, \bibinfo {author} {\bibfnamefont {Q.}~\bibnamefont {Tian}}, \bibinfo {author} {\bibfnamefont {Q.}~\bibnamefont {Deng}}, \bibinfo {author} {\bibfnamefont {D.}~\bibnamefont {Moseley}}, \bibinfo {author} {\bibfnamefont {A.}~\bibnamefont {Chen}}, \bibinfo {author} {\bibfnamefont {J.}~\bibnamefont {Lapano}}, \bibinfo {author} {\bibfnamefont {J.~S.}\ \bibnamefont {Gardner}}, \bibinfo {author} {\bibfnamefont {G.}~\bibnamefont {Eres}}, \bibinfo {author} {\bibfnamefont {T.~Z.}\ \bibnamefont {Ward}}, \bibinfo {author} {\bibfnamefont {E.}~\bibnamefont {Feng}}, \bibinfo {author} {\bibfnamefont {H.}~\bibnamefont {Cao}}, \bibinfo {author} {\bibfnamefont {V.}~\bibnamefont {Lauter}}, \bibinfo {author} {\bibfnamefont {M.~A.}\ \bibnamefont {McGuire}}, \bibinfo
  {author} {\bibfnamefont {R.}~\bibnamefont {Hermann}}, \bibinfo {author} {\bibfnamefont {D.}~\bibnamefont {Parker}}, \bibinfo {author} {\bibfnamefont {M.}~\bibnamefont {Han}}, \bibinfo {author} {\bibfnamefont {A.}~\bibnamefont {Kayani}}, \bibinfo {author} {\bibfnamefont {G.}~\bibnamefont {Rimal}}, \bibinfo {author} {\bibfnamefont {L.}~\bibnamefont {Wu}}, \bibinfo {author} {\bibfnamefont {T.~R.}\ \bibnamefont {Charlton}}, \bibinfo {author} {\bibfnamefont {R.~G.}\ \bibnamefont {Moore}},\ and\ \bibinfo {author} {\bibfnamefont {M.}~\bibnamefont {Brahlek}},\ }\bibfield  {title} {\bibinfo {title} {{Stoichiometry‐Induced Ferromagnetism in Altermagnetic Candidate MnTe}},\ }\href {https://doi.org/10.1002/adfm.202405829} {\bibfield  {journal} {\bibinfo  {journal} {Adv. Funct. Mater.}\ }\textbf {\bibinfo {volume} {34}},\ \bibinfo {pages} {2405829} (\bibinfo {year} {2024})}\BibitemShut {NoStop}%
\bibitem [{\citenamefont {Baral}\ \emph {et~al.}(2023)\citenamefont {Baral}, \citenamefont {Abeykoon}, \citenamefont {Campbell},\ and\ \citenamefont {Frandsen}}]{https://doi.org/10.1002/adfm.202305247}%
  \BibitemOpen
  \bibfield  {author} {\bibinfo {author} {\bibfnamefont {R.}~\bibnamefont {Baral}}, \bibinfo {author} {\bibfnamefont {A.~M.}\ \bibnamefont {Abeykoon}}, \bibinfo {author} {\bibfnamefont {B.~J.}\ \bibnamefont {Campbell}},\ and\ \bibinfo {author} {\bibfnamefont {B.~A.}\ \bibnamefont {Frandsen}},\ }\bibfield  {title} {\bibinfo {title} {{Giant Spontaneous Magnetostriction in MnTe Driven by a Novel Magnetostructural Coupling Mechanism}},\ }\href {https://doi.org/10.1002/adfm.202305247} {\bibfield  {journal} {\bibinfo  {journal} {Adv. Funct. Mater.}\ }\textbf {\bibinfo {volume} {33}},\ \bibinfo {pages} {2305247} (\bibinfo {year} {2023})}\BibitemShut {NoStop}%
\bibitem [{\citenamefont {Aoyama}\ and\ \citenamefont {Ohgushi}(2024)}]{PhysRevMaterials.8.L041402}%
  \BibitemOpen
  \bibfield  {author} {\bibinfo {author} {\bibfnamefont {T.}~\bibnamefont {Aoyama}}\ and\ \bibinfo {author} {\bibfnamefont {K.}~\bibnamefont {Ohgushi}},\ }\bibfield  {title} {\bibinfo {title} {{Piezomagnetic properties in altermagnetic MnTe}},\ }\href {https://doi.org/10.1103/PhysRevMaterials.8.L041402} {\bibfield  {journal} {\bibinfo  {journal} {Phys. Rev. Mater.}\ }\textbf {\bibinfo {volume} {8}},\ \bibinfo {pages} {L041402} (\bibinfo {year} {2024})}\BibitemShut {NoStop}%
\bibitem [{\citenamefont {Szuszkiewicz}\ \emph {et~al.}(2006)\citenamefont {Szuszkiewicz}, \citenamefont {Dynowska}, \citenamefont {Witkowska},\ and\ \citenamefont {Hennion}}]{PhysRevB.73.104403}%
  \BibitemOpen
  \bibfield  {author} {\bibinfo {author} {\bibfnamefont {W.}~\bibnamefont {Szuszkiewicz}}, \bibinfo {author} {\bibfnamefont {E.}~\bibnamefont {Dynowska}}, \bibinfo {author} {\bibfnamefont {B.}~\bibnamefont {Witkowska}},\ and\ \bibinfo {author} {\bibfnamefont {B.}~\bibnamefont {Hennion}},\ }\bibfield  {title} {\bibinfo {title} {{Spin-wave measurements on hexagonal $\mathrm{MnTe}$ of $\mathrm{NiAs}$-type structure by inelastic neutron scattering}},\ }\href {https://doi.org/10.1103/PhysRevB.73.104403} {\bibfield  {journal} {\bibinfo  {journal} {Phys. Rev. B}\ }\textbf {\bibinfo {volume} {73}},\ \bibinfo {pages} {104403} (\bibinfo {year} {2006})}\BibitemShut {NoStop}%
\bibitem [{\citenamefont {Kriegner}\ \emph {et~al.}(2017)\citenamefont {Kriegner}, \citenamefont {Reichlova}, \citenamefont {Grenzer}, \citenamefont {Schmidt}, \citenamefont {Ressouche}, \citenamefont {Godinho}, \citenamefont {Wagner}, \citenamefont {Martin}, \citenamefont {Shick}, \citenamefont {Volobuev}, \citenamefont {Springholz}, \citenamefont {Hol\'y}, \citenamefont {Wunderlich}, \citenamefont {Jungwirth},\ and\ \citenamefont {V\'yborn\'y}}]{PhysRevB.96.214418}%
  \BibitemOpen
  \bibfield  {author} {\bibinfo {author} {\bibfnamefont {D.}~\bibnamefont {Kriegner}}, \bibinfo {author} {\bibfnamefont {H.}~\bibnamefont {Reichlova}}, \bibinfo {author} {\bibfnamefont {J.}~\bibnamefont {Grenzer}}, \bibinfo {author} {\bibfnamefont {W.}~\bibnamefont {Schmidt}}, \bibinfo {author} {\bibfnamefont {E.}~\bibnamefont {Ressouche}}, \bibinfo {author} {\bibfnamefont {J.}~\bibnamefont {Godinho}}, \bibinfo {author} {\bibfnamefont {T.}~\bibnamefont {Wagner}}, \bibinfo {author} {\bibfnamefont {S.~Y.}\ \bibnamefont {Martin}}, \bibinfo {author} {\bibfnamefont {A.~B.}\ \bibnamefont {Shick}}, \bibinfo {author} {\bibfnamefont {V.~V.}\ \bibnamefont {Volobuev}}, \bibinfo {author} {\bibfnamefont {G.}~\bibnamefont {Springholz}}, \bibinfo {author} {\bibfnamefont {V.}~\bibnamefont {Hol\'y}}, \bibinfo {author} {\bibfnamefont {J.}~\bibnamefont {Wunderlich}}, \bibinfo {author} {\bibfnamefont {T.}~\bibnamefont {Jungwirth}},\ and\ \bibinfo {author} {\bibfnamefont {K.}~\bibnamefont {V\'yborn\'y}},\ }\bibfield  {title}
  {\bibinfo {title} {{Magnetic anisotropy in antiferromagnetic hexagonal MnTe}},\ }\href {https://doi.org/10.1103/PhysRevB.96.214418} {\bibfield  {journal} {\bibinfo  {journal} {Phys. Rev. B}\ }\textbf {\bibinfo {volume} {96}},\ \bibinfo {pages} {214418} (\bibinfo {year} {2017})}\BibitemShut {NoStop}%
\bibitem [{\citenamefont {Liu}\ \emph {et~al.}(2024)\citenamefont {Liu}, \citenamefont {Ozeki}, \citenamefont {Asai}, \citenamefont {Itoh},\ and\ \citenamefont {Masuda}}]{PhysRevLett.133.156702}%
  \BibitemOpen
  \bibfield  {author} {\bibinfo {author} {\bibfnamefont {Z.}~\bibnamefont {Liu}}, \bibinfo {author} {\bibfnamefont {M.}~\bibnamefont {Ozeki}}, \bibinfo {author} {\bibfnamefont {S.}~\bibnamefont {Asai}}, \bibinfo {author} {\bibfnamefont {S.}~\bibnamefont {Itoh}},\ and\ \bibinfo {author} {\bibfnamefont {T.}~\bibnamefont {Masuda}},\ }\bibfield  {title} {\bibinfo {title} {{Chiral Split Magnon in Altermagnetic MnTe}},\ }\href {https://doi.org/10.1103/PhysRevLett.133.156702} {\bibfield  {journal} {\bibinfo  {journal} {Phys. Rev. Lett.}\ }\textbf {\bibinfo {volume} {133}},\ \bibinfo {pages} {156702} (\bibinfo {year} {2024})}\BibitemShut {NoStop}%
\bibitem [{\citenamefont {Mazin}(2023)}]{PhysRevB.107.L100418}%
  \BibitemOpen
  \bibfield  {author} {\bibinfo {author} {\bibfnamefont {I.~I.}\ \bibnamefont {Mazin}},\ }\bibfield  {title} {\bibinfo {title} {{Altermagnetism in MnTe: Origin, predicted manifestations, and routes to detwinning}},\ }\href {https://doi.org/10.1103/PhysRevB.107.L100418} {\bibfield  {journal} {\bibinfo  {journal} {Phys. Rev. B}\ }\textbf {\bibinfo {volume} {107}},\ \bibinfo {pages} {L100418} (\bibinfo {year} {2023})}\BibitemShut {NoStop}%
\bibitem [{\citenamefont {Mu}\ \emph {et~al.}(2019)\citenamefont {Mu}, \citenamefont {Hermann}, \citenamefont {Gorsse}, \citenamefont {Zhao}, \citenamefont {Manley}, \citenamefont {Fishman},\ and\ \citenamefont {Lindsay}}]{PhysRevMaterials.3.025403}%
  \BibitemOpen
  \bibfield  {author} {\bibinfo {author} {\bibfnamefont {S.}~\bibnamefont {Mu}}, \bibinfo {author} {\bibfnamefont {R.~P.}\ \bibnamefont {Hermann}}, \bibinfo {author} {\bibfnamefont {S.}~\bibnamefont {Gorsse}}, \bibinfo {author} {\bibfnamefont {H.}~\bibnamefont {Zhao}}, \bibinfo {author} {\bibfnamefont {M.~E.}\ \bibnamefont {Manley}}, \bibinfo {author} {\bibfnamefont {R.~S.}\ \bibnamefont {Fishman}},\ and\ \bibinfo {author} {\bibfnamefont {L.}~\bibnamefont {Lindsay}},\ }\bibfield  {title} {\bibinfo {title} {{Phonons, magnons, and lattice thermal transport in antiferromagnetic semiconductor MnTe}},\ }\href {https://doi.org/10.1103/PhysRevMaterials.3.025403} {\bibfield  {journal} {\bibinfo  {journal} {Phys. Rev. Mater.}\ }\textbf {\bibinfo {volume} {3}},\ \bibinfo {pages} {025403} (\bibinfo {year} {2019})}\BibitemShut {NoStop}%
\bibitem [{\citenamefont {Rooj}\ \emph {et~al.}(2024)\citenamefont {Rooj}, \citenamefont {Chakraborty},\ and\ \citenamefont {Ganguli}}]{https://doi.org/10.1002/apxr.202300050}%
  \BibitemOpen
  \bibfield  {author} {\bibinfo {author} {\bibfnamefont {S.}~\bibnamefont {Rooj}}, \bibinfo {author} {\bibfnamefont {J.}~\bibnamefont {Chakraborty}},\ and\ \bibinfo {author} {\bibfnamefont {N.}~\bibnamefont {Ganguli}},\ }\bibfield  {title} {\bibinfo {title} {{Hexagonal MnTe with Antiferromagnetic Spin Splitting and Hidden Rashba–Dresselhaus Interaction for Antiferromagnetic Spintronics}},\ }\href {https://doi.org/10.1002/apxr.202300050} {\bibfield  {journal} {\bibinfo  {journal} {Adv. Phys. Res.}\ }\textbf {\bibinfo {volume} {3}},\ \bibinfo {pages} {2300050} (\bibinfo {year} {2024})}\BibitemShut {NoStop}%
\bibitem [{\citenamefont {Kresse}\ and\ \citenamefont {Joubert}(1999)}]{PhysRevB.59.1758}%
  \BibitemOpen
  \bibfield  {author} {\bibinfo {author} {\bibfnamefont {G.}~\bibnamefont {Kresse}}\ and\ \bibinfo {author} {\bibfnamefont {D.}~\bibnamefont {Joubert}},\ }\bibfield  {title} {\bibinfo {title} {From ultrasoft pseudopotentials to the projector augmented-wave method},\ }\href {https://doi.org/10.1103/PhysRevB.59.1758} {\bibfield  {journal} {\bibinfo  {journal} {Phys. Rev. B}\ }\textbf {\bibinfo {volume} {59}},\ \bibinfo {pages} {1758} (\bibinfo {year} {1999})}\BibitemShut {NoStop}%
\bibitem [{\citenamefont {Perdew}\ \emph {et~al.}(1996)\citenamefont {Perdew}, \citenamefont {Burke},\ and\ \citenamefont {Ernzerhof}}]{PBE}%
  \BibitemOpen
  \bibfield  {author} {\bibinfo {author} {\bibfnamefont {J.~P.}\ \bibnamefont {Perdew}}, \bibinfo {author} {\bibfnamefont {K.}~\bibnamefont {Burke}},\ and\ \bibinfo {author} {\bibfnamefont {M.}~\bibnamefont {Ernzerhof}},\ }\bibfield  {title} {\bibinfo {title} {Generalized gradient approximation made simple},\ }\href {https://doi.org/10.1103/PhysRevLett.77.3865} {\bibfield  {journal} {\bibinfo  {journal} {Phys. Rev. Lett.}\ }\textbf {\bibinfo {volume} {77}},\ \bibinfo {pages} {3865} (\bibinfo {year} {1996})}\BibitemShut {NoStop}%
\bibitem [{\citenamefont {Mosleh}\ and\ \citenamefont {Alaei}(2023)}]{Mosleh2023}%
  \BibitemOpen
  \bibfield  {author} {\bibinfo {author} {\bibfnamefont {Z.}~\bibnamefont {Mosleh}}\ and\ \bibinfo {author} {\bibfnamefont {M.}~\bibnamefont {Alaei}},\ }\bibfield  {title} {\bibinfo {title} {Benchmarking density functional theory on the prediction of antiferromagnetic transition temperatures},\ }\href {https://doi.org/10.1103/PhysRevB.108.144413} {\bibfield  {journal} {\bibinfo  {journal} {Phys. Rev. B}\ }\textbf {\bibinfo {volume} {108}},\ \bibinfo {pages} {144413} (\bibinfo {year} {2023})}\BibitemShut {NoStop}%
\bibitem [{\citenamefont {Alaei}\ and\ \citenamefont {Karimi}(2023)}]{Alaei_2023}%
  \BibitemOpen
  \bibfield  {author} {\bibinfo {author} {\bibfnamefont {M.}~\bibnamefont {Alaei}}\ and\ \bibinfo {author} {\bibfnamefont {H.}~\bibnamefont {Karimi}},\ }\bibfield  {title} {\bibinfo {title} {{A deep investigation of NiO and MnO through the first principle calculations and Monte Carlo simulations}},\ }\href {https://doi.org/10.1088/2516-1075/acbff8} {\bibfield  {journal} {\bibinfo  {journal} {Electron. Struct.}\ }\textbf {\bibinfo {volume} {5}},\ \bibinfo {pages} {025001} (\bibinfo {year} {2023})}\BibitemShut {NoStop}%
\bibitem [{\citenamefont {Szilva}\ \emph {et~al.}(2023)\citenamefont {Szilva}, \citenamefont {Kvashnin}, \citenamefont {Stepanov}, \citenamefont {Nordstr\"om}, \citenamefont {Eriksson}, \citenamefont {Lichtenstein},\ and\ \citenamefont {Katsnelson}}]{RevModPhys.95.035004}%
  \BibitemOpen
  \bibfield  {author} {\bibinfo {author} {\bibfnamefont {A.}~\bibnamefont {Szilva}}, \bibinfo {author} {\bibfnamefont {Y.}~\bibnamefont {Kvashnin}}, \bibinfo {author} {\bibfnamefont {E.~A.}\ \bibnamefont {Stepanov}}, \bibinfo {author} {\bibfnamefont {L.}~\bibnamefont {Nordstr\"om}}, \bibinfo {author} {\bibfnamefont {O.}~\bibnamefont {Eriksson}}, \bibinfo {author} {\bibfnamefont {A.~I.}\ \bibnamefont {Lichtenstein}},\ and\ \bibinfo {author} {\bibfnamefont {M.~I.}\ \bibnamefont {Katsnelson}},\ }\bibfield  {title} {\bibinfo {title} {Quantitative theory of magnetic interactions in solids},\ }\href {https://doi.org/10.1103/RevModPhys.95.035004} {\bibfield  {journal} {\bibinfo  {journal} {Rev. Mod. Phys.}\ }\textbf {\bibinfo {volume} {95}},\ \bibinfo {pages} {035004} (\bibinfo {year} {2023})}\BibitemShut {NoStop}%
\bibitem [{\citenamefont {Alaei}\ and\ \citenamefont {Oganov}(2024)}]{superhex}%
  \BibitemOpen
  \bibfield  {author} {\bibinfo {author} {\bibfnamefont {M.}~\bibnamefont {Alaei}}\ and\ \bibinfo {author} {\bibfnamefont {A.~R.}\ \bibnamefont {Oganov}},\ }\href {https://arxiv.org/abs/2410.14356} {\bibinfo {title} {{Optimizing Supercell Structures for Heisenberg Exchange Interaction Calculations}}} (\bibinfo {year} {2024}),\ \Eprint {https://arxiv.org/abs/2410.14356} {arXiv:2410.14356 [cond-mat.mtrl-sci]} \BibitemShut {NoStop}%
\bibitem [{\citenamefont {Mimasaka}\ \emph {et~al.}(1987)\citenamefont {Mimasaka}, \citenamefont {Sakamoto}, \citenamefont {Murata}, \citenamefont {Fujii},\ and\ \citenamefont {Onodera}}]{MnTe-phase}%
  \BibitemOpen
  \bibfield  {author} {\bibinfo {author} {\bibfnamefont {M.}~\bibnamefont {Mimasaka}}, \bibinfo {author} {\bibfnamefont {I.}~\bibnamefont {Sakamoto}}, \bibinfo {author} {\bibfnamefont {K.}~\bibnamefont {Murata}}, \bibinfo {author} {\bibfnamefont {Y.}~\bibnamefont {Fujii}},\ and\ \bibinfo {author} {\bibfnamefont {A.}~\bibnamefont {Onodera}},\ }\bibfield  {title} {\bibinfo {title} {Pressure-induced phase transitions of mnte},\ }\href {https://doi.org/10.1088/0022-3719/20/29/007} {\bibfield  {journal} {\bibinfo  {journal} {J. Phys. C: Solid State Phys.}\ }\textbf {\bibinfo {volume} {20}},\ \bibinfo {pages} {4689} (\bibinfo {year} {1987})}\BibitemShut {NoStop}%
\bibitem [{\citenamefont {Evans}\ \emph {et~al.}(2014)\citenamefont {Evans}, \citenamefont {Fan}, \citenamefont {Chureemart}, \citenamefont {Ostler}, \citenamefont {Ellis},\ and\ \citenamefont {Chantrell}}]{vampire}%
  \BibitemOpen
  \bibfield  {author} {\bibinfo {author} {\bibfnamefont {R.~F.~L.}\ \bibnamefont {Evans}}, \bibinfo {author} {\bibfnamefont {W.~J.}\ \bibnamefont {Fan}}, \bibinfo {author} {\bibfnamefont {P.}~\bibnamefont {Chureemart}}, \bibinfo {author} {\bibfnamefont {T.~A.}\ \bibnamefont {Ostler}}, \bibinfo {author} {\bibfnamefont {M.~O.~A.}\ \bibnamefont {Ellis}},\ and\ \bibinfo {author} {\bibfnamefont {R.~W.}\ \bibnamefont {Chantrell}},\ }\bibfield  {title} {\bibinfo {title} {{Atomistic spin model simulations of magnetic nanomaterials}},\ }\href {https://doi.org/10.1088/0953-8984/26/10/103202} {\bibfield  {journal} {\bibinfo  {journal} {J. Phys. Condens. Matter}\ }\textbf {\bibinfo {volume} {26}},\ \bibinfo {pages} {103202} (\bibinfo {year} {2014})}\BibitemShut {NoStop}%
\bibitem [{SM()}]{SM}%
  \BibitemOpen
  \href@noop {} {}\bibinfo {note} {See Supplemental Material at http://link.aps.org/supplemental/10.1103/PhysRevB.111.104416 for additional data.}\BibitemShut {Stop}%
\bibitem [{\citenamefont {Rezende}\ \emph {et~al.}(2019)\citenamefont {Rezende}, \citenamefont {Azevedo},\ and\ \citenamefont {Rodríguez-Suárez}}]{10.1063/1.5109132}%
  \BibitemOpen
  \bibfield  {author} {\bibinfo {author} {\bibfnamefont {S.~M.}\ \bibnamefont {Rezende}}, \bibinfo {author} {\bibfnamefont {A.}~\bibnamefont {Azevedo}},\ and\ \bibinfo {author} {\bibfnamefont {R.~L.}\ \bibnamefont {Rodríguez-Suárez}},\ }\bibfield  {title} {\bibinfo {title} {{Introduction to antiferromagnetic magnons}},\ }\href {https://doi.org/10.1063/1.5109132} {\bibfield  {journal} {\bibinfo  {journal} {J. Appl. Phys.}\ }\textbf {\bibinfo {volume} {126}},\ \bibinfo {pages} {151101} (\bibinfo {year} {2019})}\BibitemShut {NoStop}%
\bibitem [{\citenamefont {Yumnam}\ \emph {et~al.}(2024)\citenamefont {Yumnam}, \citenamefont {Moseley}, \citenamefont {Paddison}, \citenamefont {Suggs}, \citenamefont {Zappala}, \citenamefont {Parker}, \citenamefont {Granroth}, \citenamefont {Morris}, \citenamefont {Polash}, \citenamefont {Vashaee}, \citenamefont {McGuire}, \citenamefont {Zhao}, \citenamefont {Manley}, \citenamefont {Frandsen},\ and\ \citenamefont {Hermann}}]{PhysRevB.109.214434}%
  \BibitemOpen
  \bibfield  {author} {\bibinfo {author} {\bibfnamefont {G.}~\bibnamefont {Yumnam}}, \bibinfo {author} {\bibfnamefont {D.~H.}\ \bibnamefont {Moseley}}, \bibinfo {author} {\bibfnamefont {J.~A.~M.}\ \bibnamefont {Paddison}}, \bibinfo {author} {\bibfnamefont {C.~Z.}\ \bibnamefont {Suggs}}, \bibinfo {author} {\bibfnamefont {E.}~\bibnamefont {Zappala}}, \bibinfo {author} {\bibfnamefont {D.~S.}\ \bibnamefont {Parker}}, \bibinfo {author} {\bibfnamefont {G.~E.}\ \bibnamefont {Granroth}}, \bibinfo {author} {\bibfnamefont {G.~D.}\ \bibnamefont {Morris}}, \bibinfo {author} {\bibfnamefont {M.~M.~H.}\ \bibnamefont {Polash}}, \bibinfo {author} {\bibfnamefont {D.}~\bibnamefont {Vashaee}}, \bibinfo {author} {\bibfnamefont {M.~A.}\ \bibnamefont {McGuire}}, \bibinfo {author} {\bibfnamefont {H.}~\bibnamefont {Zhao}}, \bibinfo {author} {\bibfnamefont {M.~E.}\ \bibnamefont {Manley}}, \bibinfo {author} {\bibfnamefont {B.~A.}\ \bibnamefont {Frandsen}},\ and\ \bibinfo {author} {\bibfnamefont {R.~P.}\ \bibnamefont {Hermann}},\
  }\bibfield  {title} {\bibinfo {title} {{Magnon gap tuning in lithium-doped MnTe}},\ }\href {https://doi.org/10.1103/PhysRevB.109.214434} {\bibfield  {journal} {\bibinfo  {journal} {Phys. Rev. B}\ }\textbf {\bibinfo {volume} {109}},\ \bibinfo {pages} {214434} (\bibinfo {year} {2024})}\BibitemShut {NoStop}%
\bibitem [{\citenamefont {Mugiraneza}\ and\ \citenamefont {Hallas}(2022)}]{Mugiraneza_2022}%
  \BibitemOpen
  \bibfield  {author} {\bibinfo {author} {\bibfnamefont {S.}~\bibnamefont {Mugiraneza}}\ and\ \bibinfo {author} {\bibfnamefont {A.~M.}\ \bibnamefont {Hallas}},\ }\bibfield  {title} {\bibinfo {title} {{Tutorial: a beginner’s guide to interpreting magnetic susceptibility data with the Curie-Weiss law}},\ }\href {https://doi.org/10.1038/s42005-022-00853-y} {\bibfield  {journal} {\bibinfo  {journal} {Commun. Phys.}\ }\textbf {\bibinfo {volume} {5}},\ \bibinfo {pages} {95} (\bibinfo {year} {2022})}\BibitemShut {NoStop}%
\bibitem [{\citenamefont {Banewicz}\ \emph {et~al.}(1961)\citenamefont {Banewicz}, \citenamefont {Heidelberg},\ and\ \citenamefont {Luxem}}]{doi:10.1021/j100822a006}%
  \BibitemOpen
  \bibfield  {author} {\bibinfo {author} {\bibfnamefont {J.~J.}\ \bibnamefont {Banewicz}}, \bibinfo {author} {\bibfnamefont {R.~F.}\ \bibnamefont {Heidelberg}},\ and\ \bibinfo {author} {\bibfnamefont {A.~H.}\ \bibnamefont {Luxem}},\ }\bibfield  {title} {\bibinfo {title} {{High temperature magnetic susceptibilities of MnO, MnSe and MnTe}},\ }\href {https://doi.org/10.1021/j100822a006} {\bibfield  {journal} {\bibinfo  {journal} {J. Phys. Chem.}\ }\textbf {\bibinfo {volume} {65}},\ \bibinfo {pages} {615} (\bibinfo {year} {1961})}\BibitemShut {NoStop}%
\bibitem [{\citenamefont {Komatsubara}\ \emph {et~al.}(1963)\citenamefont {Komatsubara}, \citenamefont {Murakami},\ and\ \citenamefont {Hirahara}}]{doi:10.1143/JPSJ.18.356}%
  \BibitemOpen
  \bibfield  {author} {\bibinfo {author} {\bibfnamefont {T.}~\bibnamefont {Komatsubara}}, \bibinfo {author} {\bibfnamefont {M.}~\bibnamefont {Murakami}},\ and\ \bibinfo {author} {\bibfnamefont {E.}~\bibnamefont {Hirahara}},\ }\bibfield  {title} {\bibinfo {title} {{Magnetic Properties of Manganese Telluride Single Crystals}},\ }\href {https://doi.org/10.1143/JPSJ.18.356} {\bibfield  {journal} {\bibinfo  {journal} {J. Phys. Soc. Jpn.}\ }\textbf {\bibinfo {volume} {18}},\ \bibinfo {pages} {356} (\bibinfo {year} {1963})}\BibitemShut {NoStop}%
\end{thebibliography}%

%\newpage
%\onecolumngrid
%\appendix
%\renewcommand{\thefigure}{S\arabic{figure}}
%\setcounter{figure}{0}  

\end{document}